\documentclass[prd,preprint,tightenlines,floatfix,showpacs,showkeys,preprintnumbers,nofootinbib,eqsecnum]{revtex4}
 \usepackage[dvips,final]{graphicx}
  \usepackage{bm}
   \usepackage{amsmath}
    \usepackage{amssymb}
     \usepackage{pifont}

\newcommand{\footn}{\footnotesize}                        
 \newcommand{\LI}[1]{\textbf{Li}_{#1}}                    
  \newcommand{\Ds}{\displaystyle}                         
   \newcommand{\va}[1]{\langle{#1}\rangle}                
    \newcommand\convo[1]{\mathop{\otimes}\limits_{#1}}    
\newcommand{\gev}[1]{\relax\ifmmode{\text{GeV}^{#1}}      
                     \else{GeV$^{#1}${ }}\fi}             
\newcommand{\Gev}{\relax\ifmmode{\text{GeV}}              
                     \else{GeV{ }}\fi}                    
\newcommand{\Mev}{\relax\ifmmode{\text{MeV}}              
                     \else{MeV{ }}\fi}                    
\def\as{\relax\ifmmode\alpha_s\else{$ \alpha_s${ }}\fi}   
\newlength{\tabcolf} 
\newlength{\tabcols} 
\newlength{\tabcolt} 

\begin{document}
\thispagestyle{empty}

\preprint{RUB-TPII-04, JINR-E2-2002-278}
\title{Unbiased analysis of CLEO data at NLO
        and pion distribution amplitude}

\author{Alexander~P.~Bakulev}%
 \email{bakulev@thsun1.jinr.ru}
\author{S.~V.~Mikhailov}%
 \email{mikhs@thsun1.jinr.ru}
\affiliation{%
  Bogoliubov Laboratory of Theoretical Physics,  \\
  Joint Institute for Nuclear Research,          \\
  141980, Moscow Region, Dubna, Russia}%

\author{N.~G.~Stefanis}
 \email{stefanis@tp2.ruhr-uni-bochum.de}
\affiliation{%
  Institut f\"ur Theoretische Physik II          \\
  Ruhr-Universit\"at Bochum                      \\
  D-44780 Bochum, Germany}%
\vspace {10mm}

\begin{abstract}
We discuss different QCD approaches to calculate the form factor
$F^{\gamma^*\gamma\pi}(Q^2)$ of the $\gamma^*\gamma\to\pi^{0}$
transition giving preference to the light-cone QCD sum rules (LCSR)
approach as being the most adequate.
In this context we revise the previous analysis of the CLEO
experimental data on
$F^{\gamma^*\gamma\pi}\left(Q^{2}\right)$
by Schmedding and Yakovlev. Special attention is paid to the
sensitivity of the results to the (strong radiative)
$\alpha_s$-corrections and to the value of the twist-four coupling
$\delta^2$. We present a full analysis of the CLEO data at the NLO
level of LCSRs, focusing particular attention to the extraction of
the relevant parameters to determine the pion distribution amplitude,
i.e., the Gegenbauer coefficients $a_2,~a_4$. Our analysis confirms
our previous results and also the main findings of Schmedding and
Yakovlev: both the asymptotic, as well as the Chernyak--Zhitnitsky
pion distribution amplitudes are completely excluded by the CLEO
data. A novelty of our approach is to use the CLEO data as a means
of determining the value of the QCD vacuum non-locality parameter
$\lambda^2_q=\va{\bar{q}D^2q}/\va{\bar{q}q}=0.4\text{ GeV}^2$,
which specifies the average virtuality of the vacuum quarks.
\end{abstract}
\vspace {2mm}

\maketitle
\newpage
\vspace*{10mm}

\section{Introduction}
Recently, the CLEO collaboration \cite{CLEO98} has measured the
$\gamma^{*}\gamma \to \pi^{0}$ form factor
$F^{\gamma^{*}\gamma\pi}(Q^2)$ with high precision.
This data has been processed by Schmedding and Yakovlev
(SY) \cite{SchmYa99} using light-cone QCD sum rules (LCSR),
taking also into account the perturbative QCD contributions
in the next-to-leading order (NLO) approximation.
In this way SY obtained useful constraints on the shape of the pion
distribution amplitude (DA) in terms of confidence regions for the
Gegenbauer coefficients $a_{2}$ and $a_{4}$, the latter being the
projection coefficients of the pion DA on the corresponding
eigenfunctions.
Note that SY have extended to the NLO the LCSR approach suggested
before by Khodjamirian \cite{Kho99} for the leading order (LO)
light-cone sum rule method.

The present analysis gives further support to the claim,
expressed by the above mentioned authors,
that LCSRs provide the most appropriate basis
in describing the form factor
of the $\gamma^*\gamma\to\pi^0$ transition.
This is intimately connected with peculiarities
of real-photon processes in QCD \cite{RR96,Kho99}.
But the method of the CLEO data processing, adopted in~\cite{SchmYa99},
seems to be not quite complete from our point of view.
We think that an optimal analysis should take into account the correct
ERBL evolution of the pion DA to the scale $Q^2_\text{exp}$ of the process
(the latter not to be fixed at some average point, $\mu_\text{SY}=2.4$~GeV,
as done in \cite{SchmYa99}) and to re-estimate the contribution $\delta^2$
from the next twist term.
The influence of both these effects appears to be important and it is
examined here in detail.
Furthermore, we are not satisfied with the error estimation performed in the
SY analysis, for reasons to be explained later, and prefer therefore to use
a more traditional treatment to determine the sensitivity to the input parameter
$\delta^2$ and the construction of the 1-$\sigma$ and 2-$\sigma$ error contours.

Our main goal in the present work will be to obtain new constraints on the
$(a_{2},~a_{4})$ DA parameters from the CLEO data, taking into account all
the remarks mentioned above, and then to compare them with the constraints
following from QCD SRs with nonlocal condensates (NLC).
We will not repeat here the derivation of the main results of LCSRs,
as well as those related to NLC SRs,
but we will refer the interested reader to \cite{Kho99,SchmYa99}
and correspondingly to \cite{BMS01,BM02} and
references therein.
But for the sake of convenience we included a technical exposition of
our approach in comprehensive appendices.

The paper is organized as follows.
In Sec.\ \ref{sec:transff} we review different QCD approaches to calculate
the transition form factor $F^{\gamma^{*}\gamma\pi}(Q^2)$, having recourse
to QCD ``factorization theorems'' \cite{ERBL79},
encompassing both perturbation theory and LCSRs.
The analysis of the CLEO data is discussed in Sec.\ \ref{sec:CLEO} in
conjunction with the SY approach in comparison with other
approaches/approximations.
In Sec.\ \ref{sec:2loop} we present a complete NLO analysis of the
CLEO data with a short discussion of the BLM setting procedure.
Sec.\ \ref{sec:pionDA} includes a comparison of the QCD SR pion DA models
with the results obtained in Sec.\ \ref{sec:2loop} from the CLEO data
processing.
In Sec.\ \ref{sec:concl} we summarize our conclusions.
The paper ends with five appendices: in Appendix A we re-estimate the value
of the twist-four scale $\delta^2$.
In Appendix B the old Chernyak--Zhitnitsky (CZ) result for $a_2$ is discussed,
paying attention to evolution effects.
In Appendices C and D the two-loop results \cite{DaCh81,KMR86}
for the purely perturbative part of the form-factor calculations
and the ERBL evolution of the pion DA are outlined.
Finally, in Appendix E, all needed calculation details for the NLO LCSR
are presented.
\section{Transition form factor
  $\mathbf{F^{\gamma^{*}\gamma \pi}(Q^2)}$ with LCSR}
\label{sec:transff}
\subsection{Factorization of the $\mathbf{F^{\gamma^{*}\gamma^{*} \pi}}$
form factor. Standard results}
\label{subsec:2.1}
The form factor  of the process
$\gamma^*(q_1)\gamma^*(q_2)\to\pi^0(p)$
is defined by the matrix element
\begin{equation}
 \label{eq:ampl}
  \int d^4z e^{-iq_1z}
   \va{\pi^0(p)\mid T\{j_\mu(z) j_\nu(0)\}\mid 0}
  = i\epsilon_{\mu\nu\alpha\beta}
    q_1^\alpha q_2^\beta F^{\gamma^*\gamma^* \pi}(Q^2,q^2)\,,
\end{equation}
where $q_1 + q_2 = p$, $Q^2=-q_1^2 > 0$ with $q^2=-q_2^2\geq 0$
are the virtualities of the photons,
$j_\mu=
 (\frac23\bar{u}\gamma_\mu u
 - \frac13\bar{d}\gamma_\mu d)$
is the quark electromagnetic current, and $\pi^0(p)$ is the pion
state with momentum $p$.
If both virtualities, $Q^2$ and $q^2$, are sufficiently large, the
$T$-product of the currents in (\ref{eq:ampl}) can be expanded near the
light-cone $z^2=0$.
This expansion results in factorization theorems, which control the
structure of the form factor \cite{ERBL79} at leading twist.
As a result, the form factor can be cast in the form of a convolution
over the momentum fraction variable $x$ (of the total momentum $p$)
to read
\begin{eqnarray}
 \label{eq:factor}
  F_\text{QCD}^{\gamma^*\gamma^* \pi}(Q^2,q^2)
   = f_{\pi}
   \int\limits_0^1\!\!dx\,\,
    T(Q^2,q^2;\mu_\text{F}^2;x)
    \varphi_\pi (x;\mu_\text{F}^2)
   \equiv f_{\pi}
    T(Q^2,q^2;\mu_\text{F}^2;x)
    \convo{x}
     \varphi_\pi (x;\mu_\text{F}^2)\, .
\end{eqnarray}
The hard amplitude $T(Q^2,q^2;x)$ -- playing the role of the Wilson
coefficient in the OPE -- is calculable within QCD perturbation theory
(PT):
\begin{equation}
 \label{eq:T-pert}
  T = T_0
    + \frac{\as(\mu_\text{R}^2)}{4\pi} T_1
    + \left[\frac{\as(\mu_\text{R}^2)}{4\pi}\right]^2 T_2
    + \ldots\, ,
\end{equation}
while the pion DA $\varphi_\pi(x;\mu_\text{F}^2)$ contains the
long-distance effects and demands the application of nonperturbative
methods.
Due to factorization theorems, DAs enter as the central input of
various QCD calculations of hard exclusive processes.
The LO term $T_0$ in (\ref{eq:T-pert}) depends only on kinematical
variables, but the NLO amplitude $T_1$,
calculated in \cite{DaCh81,KMR86},
depends also on the factorization scale $\mu_\text{F}^2$:
\begin{eqnarray} \label{eq:factor0}
  T_0(Q^2,q^2;x)
  &\equiv& N_T
    C_0(Q^2,q^2;x)\
   =\ N_T
    \left[\frac{1}{Q^2x+q^2(1-x)}
         + (x \to 1-x)
    \right]\,;\\
  T_1(Q^2,q^2;\mu_\text{F}^2; x)
  &=& N_T
    \left[\ln\left(\frac{\bar{Q}^2}{\mu_\text{F}^2}\right)
           \cdot C_0(Q^2,q^2;u)\convo{u} V_0(u,x)
         + C_1(Q^2,q^2;x)
    \right]\,.
  \label{eq:factor1}
\end{eqnarray}
Here, $N_T \equiv (e_u^2-e_d^2)/\sqrt{2} = \sqrt{2}/6$ is
the QCD normalization factor,
$\bar{Q}^2=-\left[(q_1-q_2)/2\right]^2=(Q^2+q^2)/2$,
and $\mu_\text{R}^2$ and $\mu_\text{F}^2$
denote, respectively,
the scale of renormalization of the theory
and the factorization scale of the
process.
$C_{0,1}$ and $V_{0,1}$ are, respectively,
the perturbative expansion elements
of the coefficient function $C$ of the process and the ERBL kernel $V$
in the LO (subscript 0) and NLO (subscript 1) approximation
\footnote{wherein $V_1$ is needed to evolve the pion DA in the NLO
approximation.}.
The structure of $T_1$ is discussed in more detail in \cite{KMR86,MuR97}
and in Appendix \ref{App-RC}.1.
Here we only recall those features relevant for our discussion:
Eq. (\ref{eq:factor1}), which specifies the definition of the
coefficient function $C_{1}$ is a consequence of the QCD factorization
theorems \cite{ERBL79}.
Due to these theorems, the first logarithmic term in Eq.\ (\ref{eq:factor1}),
originating from collinear divergences,
can be absorbed into the renormalization of the DA,
$\varphi_\pi(x;\mu_\text{F}^2)$,
following the ERBL equation (see App.\ \ref{App-EvoDA}, Eq.\ (\ref{eq:App-ERBL})).
Namely, the formal solution of the ERBL equation in the 1-loop
approximation is
\begin{eqnarray}
 \label{eq:EvPDA}
  \varphi_\pi(x;\mu^2_0)
   \stackrel{\text{ERBL}}{\longrightarrow}
    \varphi_\pi^\text{RG}(x;\mu^2)
  &=&\exp
      \left[\int_{\alpha_s(\mu_0^2)}^{\alpha_s(\mu^2)}
       d a \frac{aV_0\ }{\beta(a)}
        \convo{}
      \right]
      \varphi_\pi(x;\mu_0^2)
\end{eqnarray}
(here $\beta(a)=-b_0 a^2$ is the 1-loop $\beta$-function)\footnote{%
and one should mean
$\Big[V_0\convo{}\Big]^{n}\varphi_\pi(x;\mu_0^2)
= V_0(x,u_1)\ \convo{u_1}\
   \ldots\convo{u_{n-1}}
    V_0(u_{n-1},u_n)\convo{u_n}
     \varphi_\pi(u_n;\mu_0^2)$.}.
We adopt here as a factorization scale $\mu_\text{F}^2=Q^2$.
For that case, $F^{\gamma^*\gamma^* \pi}$,
obtained in Eq.\ (\ref{eq:factor}) at the scale $\mu^2$,
can be transformed into
\begin{equation}
 \label{eq:factor2}
  F^{\gamma^*\gamma^* \pi} \to
  F^{\gamma^*\gamma^* \pi}_\text{RG}(Q^2,q^2) =
  f_{\pi}
  \left[T_0(Q^2,q^2;x)
        + \frac{\alpha_s(\mu_\text{R}^2)}{4\pi} T_1(Q^2,q^2;Q^2;x)
  \right]
   \convo{x}
    \varphi_\pi^\text{RG}(x;Q^2)\,.
\end{equation}
To fix the renormalization scale $\mu_\text{R}^2$, one needs to go
beyond the NLO approximation.
In the absence of such information and in order to further
simplify the NLO analysis, we also set
$\mu_\text{R}^2=\mu_\text{F}^2=Q^2$.
It is useful to expand $\varphi_\pi^\text{RG}(x;\mu^2)$ over the
eigenfunctions $\psi_n(x)$ of the one-loop ERBL equation, i.e.,
in terms of the Gegenbauer polynomials $C^{3/2}_n(\xi)$,
\begin{equation}
 \label{eq:Gegenbauer}
  \varphi_\pi^{RG}(x;\mu^2)
  =  \sum_{n=0,2,4,\ldots}a_n(\mu^2)\cdot \psi_n(x)\, ;
   \quad
    a_0=1;~\psi_n(x)
     \equiv 6x(1-x)C^{3/2}_n(2x-1)\, .
\end{equation}
In this representation, all the dependence on $\mu^2$ is contained
in the coefficients $a_n(\mu^2)$ and is dictated by the ERBL equation.
Different reasons, explained in \cite{SchmYa99} and \cite{BMS01,BM02},
point to the possibility of retaining only the first 3 terms in this
expansion.
Following then this approximation, the form factor can be parameterized
by only two variables $a_2(\mu_0^2)$ and $a_4(\mu_0^2)$ that accumulate
the mesonic large-distance effects (at some scale $\mu_0^2$).

A special case of the OPE appears when one of the photons ($q_2$) is
near the mass shell.
It is instructive to present here an evident NLO perturbative expression
for Eq.\ (\ref{eq:factor2}) in the \textit{formal limit} $q^2 \to 0$
(cf.\ Eq.\ (\ref{eq:AppB_NLO+tw4}) in Appendix \ref{App-RC}),
\begin{eqnarray}
 \label{eq:NLO}
   F_\text{PT}^{\gamma^*\gamma\pi}(Q^2)
  &=& \frac{\sqrt{2}}{Q^2}f_{\pi}
      \left\{1 + \Sigma(Q^2)
              + \frac{\alpha_s(Q^2)}{4\pi}C_\text{F}
                \left[-5 + 7.9\cdot \Sigma(Q^2) - 3.8\cdot \Delta(Q^2)
                \right]\right.
\nonumber \\
&& \qquad\qquad             \left. -\frac{80}{27}\frac{\delta^{2}(Q^2)}{Q^2}
      \right\}
\end{eqnarray}
with
$
  \Sigma(Q^2) \equiv a_2(Q^2) + a_4(Q^2)
$
and
$
  \Delta(Q^2) \equiv a_2(Q^2) - a_4(Q^2).
$

On the r.h.s. of Eq.\ (\ref{eq:NLO}), the twist-four contribution is included
in terms of the twist-four asymptotic DAs, presented in \cite{Kho99}.
The coefficients $a_{2,4}(Q^2)$ and the twist-four coupling $\delta^{2}(Q^2)$
are evolved to the scale $Q^2$ using, correspondingly, the renormalization
group (RG) equation to NLO and LO accuracy\footnote{%
The reasons for this treatment will be discussed in more detail in
Sec.\ \ref{sec:2loop}.}.

\subsection{Why do we need LCSRs for the transition form factor?}
\label{subsec:2.2}
A straightforward calculation of $F^{\gamma^*\gamma\pi}(Q^2)$
in QCD is not possible.
In particular, at the formal limit $q^2\to 0$, it is not sufficient to
retain only a few terms of the light-cone OPE of (\ref{eq:ampl}).
One has, in addition, to take into account the interaction of the
small-virtuality photon at large distances proportional to
$O(1/\sqrt{q^2})$, (for a recent discussion, consider section 4 in
\cite{RR96} and \cite{MuR97}).
The LCSR method allows one to avoid the problem of the photon
long-distance interaction by providing the means of performing all
QCD calculations at sufficiently large $q^2$ and then use a dispersion
relation to return to the mass-shell photon.
To isolate the dangerous neighborhood of $q^2=0$,
one should apply an appropriate realistic model
for the spectral density at low $s$,
based, for instance, on quark-hadron duality.
Using this sort of analysis and by employing analyticity and duality arguments,
the following expression was obtained in \cite{Kho99}
\begin{equation}
 \label{eq:srggpi}
 F_\text{LCSR}^{\gamma^*\gamma\pi}(Q^2)
 = \int\limits_0^{s_0}\!\!\frac{ds}{m_\rho^2}\,
    \mathbf{Im}
     \left[\frac1\pi\,F_\text{QCD}^{\gamma^*\gamma^*\pi}(Q^2,s)
     \right]\,
      e^{\left(m_\rho^2-s\right)/M^2}
 + \int\limits_{s_0}^\infty\!\!\frac{ds}{s}\,
    \mathbf{Im}
     \left[\frac1\pi\,F_\text{QCD}^{\gamma^*\gamma^*\pi}(Q^2,s)
     \right]\, .
\end{equation}
$F_\text{QCD}^{\gamma^*\gamma^*\pi}(Q^2,s)$
on the r.h.s. of (\ref{eq:srggpi}) is taken from Eq.\ (\ref{eq:factor})
and the Borel parameter is $M^2\approx0.7$~GeV$^2$, whereas $m_\rho$ is the
$\rho$-meson mass and $s_0=1.5$~GeV${}^2$ denotes the effective threshold
in the $\rho$-meson channel.

This program has first been suggested in general form in \cite{Kho99}
and was realized there for the LO approximation of the process.
Taking
$F_\text{QCD}^{\gamma^*\gamma^*\pi}$
at LO in $\as$, the form factor
$F_\text{LOLC}^{\gamma^*\gamma\pi}(Q^2)$ was obtained
from Eq.\ (\ref{eq:srggpi}) with
\begin{equation}
 \label{eq:ff-tw4}
 F_\text{LO QCD}^{\gamma^*\gamma^*\pi}(Q^2,q^2)
  =  f_{\pi}N_T
     \left\{C_0(Q^2,q^2;u) \convo{u} \varphi_\pi(u)
          - \frac{1}{2}\left[C_0(Q^2,q^2;u)\right]^2
             \convo{u} \varphi^{(4)}_\pi(u)
      \right\}\,
\end{equation}
and
\begin{equation}
 \mathbf{Im}
    \left[\frac1\pi\,F_\text{LO QCD}^{\gamma^*\gamma^*\pi}(Q^2,s)
    \right]
  = \frac{f_{\pi}\sqrt{2}}{3}
    \left[\frac{\varphi_\pi(u)}{s+Q^2}
         -\frac1{Q^2}\frac{d\varphi^{(4)}(u)}{ds}
    \right]_{u=Q^2/(s+Q^2)}\, .
 \label{eq:rhoqcd}
\end{equation}
Here, the numerically important twist-four contribution
was also included
using a simple asymptotic expression for the twist-four DA contribution
\cite{Kho99}
\begin{equation}
 \varphi^{(4)}(u,\mu^2)
 = \frac{80}3 \delta^2(\mu^2) u^2(1-u)^2\,.
 \label{eq:phi4}
\end{equation}
Note that the coupling $\delta^2$ is determined from the matrix element
\begin{equation}
 \va{\pi(p) |g_s\bar{d}\tilde{G}_{\alpha\mu}\gamma^\alpha u|0}
 = i\delta^2f_\pi p_\mu\, .
 \label{eq:delta}
\end{equation}
The estimates for $\delta^2(\mu^2)$, including also its RG-evolution,
are presented in Appendix \ref{App-delta}.

\subsection{The framework of NLO LCSR for the  form factor}
\label{subsec:2.4}
An application of the above scheme to the NLO was more recently performed
in \cite{SchmYa99}.
The $\as$-corrections to this process are expected to be rather large,
of the order of 20\% (see, e.g., \cite{MuR97}).
The size of these corrections can be roughly estimated
from the perturbative expressions for their asymptotic parts
by comparing to each other
the values $1$ and $\Ds -5 \frac{\as}{4\pi} C_\text{F}$
in Eq.\ (\ref{eq:NLO}).
Therefore, for a quantitative description of the form factor, this
contribution should be taken into account.
This important step was done by Schmedding and Yakovlev, who have used the
NLO perturbative expression for
$F_\text{QCD}^{\gamma^*\gamma^* \pi}$
in Eq.\ (\ref{eq:srggpi}),
\begin{eqnarray}
 \label{eq:SY}
  F_\text{NLO QCD}^{\gamma^*\gamma^*\pi}(Q^2,q^2;\mu^2)
  =
  F_\text{LO QCD}^{\gamma^*\gamma^*\pi}(Q^2,q^2)
  + \frac{\alpha_s(\mu^2)}{4\pi}
    f_{\pi} T_1(Q^2,q^2;\mu^2;x)
     \convo{x}\varphi_\pi(x;\mu^2)\,,
\end{eqnarray}
where $\mu^2\equiv \mu_\text{F}^2=\mu_\text{R}^2 \neq Q^2$,
to construct a NLO version of the LCSR for the form factor
$F^{\gamma^*\gamma\pi}(Q^2)$.
The spectral density
$\mathbf{Im}[F_\text{NLO QCD}^{\gamma^*\gamma^*\pi}(Q^2,s;\mu^2)]/\pi$,
based on Eq.\ (\ref{eq:SY}),
depends on the scale $\mu^2$.
In the original paper \cite{SchmYa99}, this scale was fixed by relating
it to the mean value of $Q^2$
with respect to the CLEO experimental data,
i.e.,
by setting $\mu^2 = \mu_\text{SY}^2 =$
$\va{Q_\text{exp}^2} \approx (2.4)^2~\gev{2}$.
Use of Eq.\ (\ref{eq:SY}) implies that one accounts for the scale
dependence in $\varphi_\pi(x;Q^2)$ (e.g., for the different CLEO points)
via the leading order perturbative formula
$$\varphi_\pi(x;Q^2)
  \approx
  \left[1
      + \frac{\alpha_s(\mu^2)}{4\pi}
         \ln\left(Q^2/\mu^2\right)\cdot V_0(x,u)
  \right]
   \convo{u}\varphi_\pi(u;\mu^2)\, ,
$$
rather, than using the RG expression, given by Eq.\ (\ref{eq:EvPDA}).
This seems to be a rather crude approximation, given that other
reference points are evolved with $\mu^2$ by utilizing the NLO ERBL
evolution equation.

\section{CLEO data analysis revisited}
\label{sec:CLEO}
Let us scrutinize the form factor approaches, discussed above,
in close comparison with the CLEO data \cite{CLEO98}.

\subsection{Theoretical approaches to the CLEO data}
\label{subsec:3.1}
In this subsection -- in order to be in close analogy with the original
SY paper -- we shall adopt their scale definition of the CZ DA.
It is worth noting, however, that the genuine CZ DA differs from that
definition because it is determined at the much lower normalization scale
$\mu_0=0.5$~GeV (a discussion of this important point is relegated to
Appendix \ref{App-CZDA}).
To distinguish the two models in the present analysis, we will use in what
follows a special notation:
(i) CZ DA parameters originating from the SY prescription
($a_2^\text{CZ}(\mu^2=0.5~\gev{2}) = 2/3$),
and by using a 2-loop evolution to the scale $\mu^2_\text{SY}$,
will be denoted with the superscripts CZSY,
whereas (ii) those conforming with our prescription
($a_2^\text{CZ}(\mu^2=1~\gev{2}) = 0.56$,
see Appendix \ref{App-CZDA})
and being 2-loop evolved to the scale $\mu^2_\text{SY}$
will be marked by the superscripts CZ.
\begin{figure}[h]
 \centerline{\includegraphics[width=0.45\textwidth]{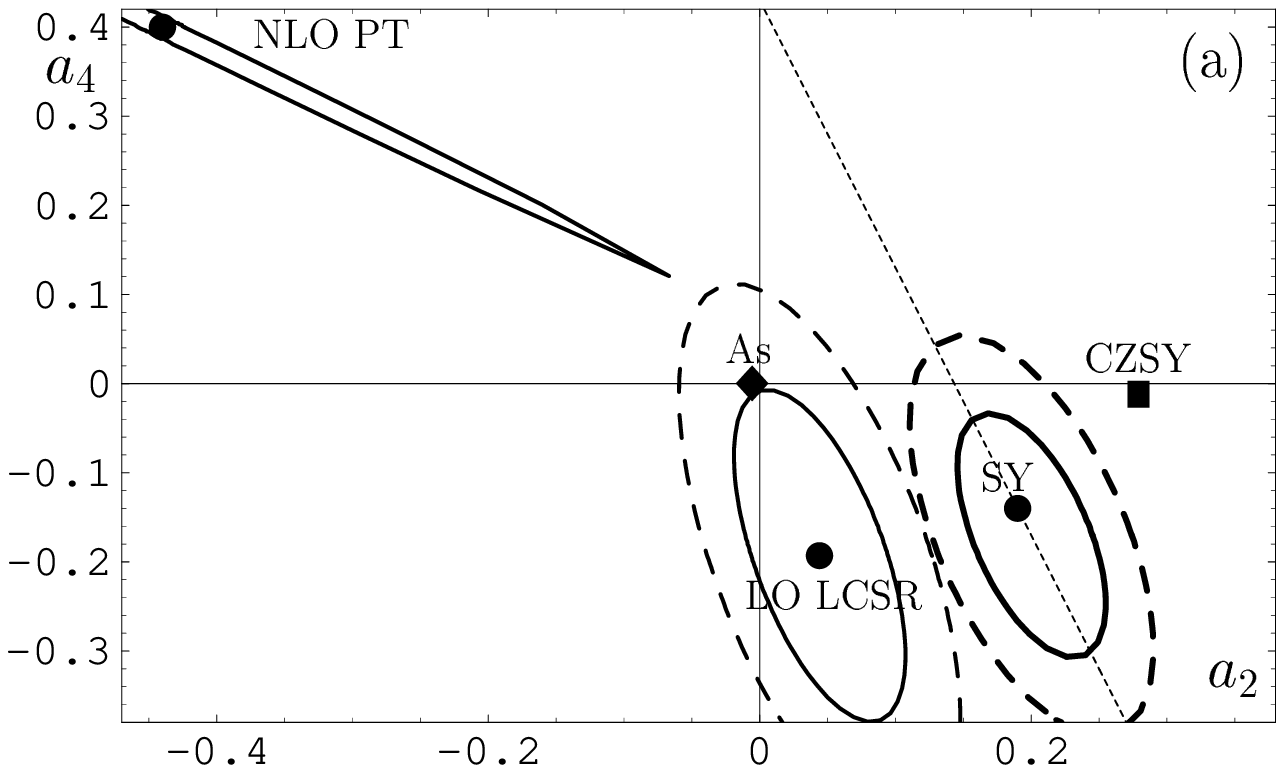}
  \hspace{0.05\textwidth}
   \includegraphics[width=0.45\textwidth]{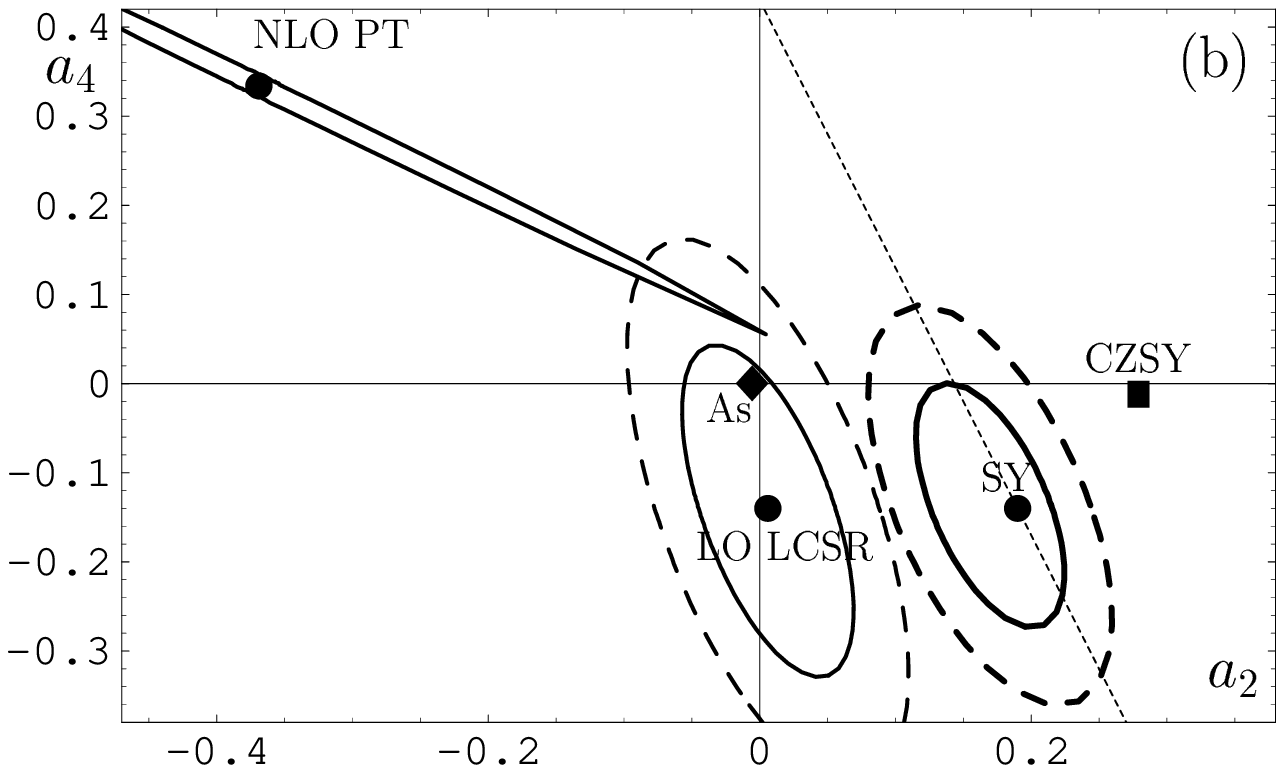}}
    \caption{\label{fig:3-ogur}
     \footn The plane ($a_2,a_4$) and results of different analyses
      of the CLEO data for the form factor $F_{\pi\gamma\gamma}(Q^2)$.
      We show here the best-fit points of LO LCSR, NLO LC SR (SY)
      and a NLO PT analysis for two values of the twist-four coupling
      $\delta^2=0.2~\gev{2}$ (a) and $\delta^2=0.18~\gev{2}$ (b).
      The solid contours correspond to a $1\sigma$-criterion,
      while the broken ones to a $2\sigma$-criterion.
      We also plot for comparison the positions of the CZSY
      ({\tiny\ding{110}}) and the asymptotic (\ding{117}) DAs.
      All values are evaluated at $\mu^2=(2.4~\gev{})^2$.
      The inclined straight line on the right hand-side
      of the figure is the modified diagonal
      $a_2+a_4/3=$\textsc{const}, corresponding to the new
      $1\sigma$-ellipse of the NLO LCSR analysis.}
\end{figure}

The status of these approximations (NLO PT, LO LCSR and NLO LCSR)
is illustrated in Fig.\ \ref{fig:3-ogur} by the relative positions of
the corresponding admissible regions for these parameters in the
$(a_2,a_4)$ plane.
Here, the regions enclosed by the needle-like- and ellipse-like solid
contours correspond to a $1\sigma$-deviation criterion
(CL=68\%)~\cite{PDG2000},
while the broken contours refer to a $2\sigma$-deviation criterion
(CL=95\%).
Note that these contours have been derived by taking into account only
the statistical error bars in~\cite{CLEO98}
(see their Table 1).
This marks a crucial difference between our processing of the data
and that in \cite{SchmYa99}, where the ``theoretical-systematic
uncertainties'' have been involved in the statistical analysis together
with the statistical ones.
In other words, we do not ``smear'' the quantities $\delta^2$
(or $a_0=1$)
over their corresponding (theoretical) error-bar intervals.
In our opinion,
such a manner would require an additional substantiation
and further suggestions about the distribution of these errors
that we want to avoid in our analysis.
Hence, instead of that, we process the data at a few fixed values of
$\delta^2$ in order to clarify
the sensitivity of the results to this parameter.

It should be clear that a really admissible region might be somewhat
larger than the presented ``purely statistical'' contours in
Fig.\ \ref{fig:3-ogur}, a price one has to pay for our strict way of
data processing.
\begin{enumerate}
\item[(i)] The large contour, enclosing the origin, corresponds to the
LO LCSR form factor, discussed in \cite{Kho99}.
Note that the point denoting the position of the asymptotic DA lies just
on the periphery of the $1\sigma$-contour and that the CZ DA one (see
\cite{Kho99}), lies far outside.
\item[(ii)] The needle-like contour on the left top corner of both parts
of Fig.\ \ref{fig:3-ogur} corresponds to Eq.\ (\ref{eq:NLO}) and is
stretched along the ``diagonal'' $a_2+0.75\cdot a_4 =$ \textsc{const}.
The weak dependence of Eq.\ (\ref{eq:NLO}) on $\Delta \equiv a_2-a_4$
slightly turns the angle of the diagonal $\simeq3\pi/4\to 0.8\pi$.
On the other hand, taking into account the evolution with $Q^2$ of $a_{2}$
and $a_{4}$ for every $Q^2_\text{exp}$ makes the contour finite --
much like a ``diagonal'' needle-like strip.
As we have mentioned above, the formal limit of the NLO expression,
Eq.\ (\ref{eq:NLO}), \textsl{cannot give} a reliable result for the form
factor.
In this context it is interesting to mention that the corresponding
contour is located outside the regions determined by the LCSRs.
Furthermore, all known DA models (see, e.g., Table~\ref{tab-1}) and the
phenomenological predictions (\cite{BKM00,BMS01} and references therein)
are located far away from this contour -- clearly demonstrating
the poor reliability of the corresponding perturbative approach.
\item[(iii)] At least the heavy-line contours 
(enclosing also the SY point \cite{SchmYa99}) 
correspond to the SY approximation.
These contours do not overlap with those corresponding to the LO LCSR
ones -- even at the $2\sigma$-deviation level.
Therefore, $\as$-corrections are crucially important in extracting
the DA parameters.
Our best-fit point with respect to the NLO LCSR is close to but not
coinciding with the one presented by SY (compare entries 3 and 4 in
Table \ref{tab-1}).
In Fig.\ \ref{fig:3-ogur}(a) the full circle inside the contour is just
the SY point.
The best-fit points, corresponding to different approximations and
models, considered in the present analysis, are collected in Table
\ref{tab-1}, where the notation $\chi^2_\text{(d.o.f.)}=\chi^2/14$ has
been used (in correspondence to the number 15 of the CLEO experimental
data points).
\end{enumerate}
\begin{table}[h]
\caption{Compilation of best-fit points shown in Fig.\ \ref{fig:3-ogur}(a)
         with $\delta^2(1~\gev{2})=0.20$~GeV${}^2$.}
\begin{center}
\begin{tabular}{|l|c|c|c|}
\colrule
  ~Best-fit points/models $\vphantom{^|_|}$
         & ~$a_2(\mu_\text{SY}^2)$~ & ~$a_4(\mu_\text{SY}^2)$~ & ~$\chi^2_\text{(d.o.f.)}$~
 \\ \hline\hline
 ~NLO PT (\ref{eq:NLO}) best fit $\vphantom{^|_|}$
         & $-0.44$             & $+0.40$                  & $0.49$
 \\ \hline
 ~LO LCSR best fit $\vphantom{^|_|}$
         & $+0.04$             & $-0.19$                  & $0.48$
 \\ \hline
 ~NLO LCSR best fit $\vphantom{^|_|}$
         & $+0.20$             &  $-0.17$                 & $0.48$
 \\ \hline \hline
 ~SY LCSR~\cite{SchmYa99}$\vphantom{^|_|}$
         & $+0.19$             & $-0.14$                  & $0.49$
 \\ \hline
 ~BMS model~\cite{BMS01} $\vphantom{^|_|}$
         & $+0.13$             & $-0.08$                  & $0.74$
 \\ \hline
 ~\text{CZSY DA~\cite{SchmYa99}} $\vphantom{^|_|}$
         & $+0.28$             & $-0.009$                  & $1.8$
 \\ \hline
 ~\text{CZ DA~\cite{CZ84}} $\vphantom{^|_|}$
         & $+0.35$             & $-0.006$                 & $3.6$
 \\ \hline 
 ~Asympt. DA $\vphantom{^|_|}$
         & $-0.006$            & $+0.00$                  & $3.0$
 \\ \hline
\end{tabular}
\label{tab-1}
\end{center}
\end{table}
It should be stressed that the admissible region (heavy-line contours),
obtained with our data-processing procedure,
differs from that in \cite{SchmYa99}.
Our contours look slightly larger than theirs despite the fact that
possible theoretical/systematic uncertainties were not included in our
consideration.
Just because of this latter reason, our contours possess another
orientation relative to those of SY,
as one appreciates by comparing
Fig.\ \ref{fig:3-ogur}(a) and Fig.\ \ref{fig:ogur-new09} with
Fig.\ 6 in \cite{SchmYa99}.
Moreover, the CZSY DA model appears to be \textsl{seemingly} closer
($\chi^2=1.8$) to the best-fit point $(0.20,-0.17)$ than the
asymptotic one ($\chi^2\approx 3$).
But the genuine CZ DA with $a_{2}^\text{CZ}(\mu_\text{SY}^2)=0.35$
and $a_{4}^\text{CZ}(\mu_\text{SY}^2)=-0.006$
(consult the discussion in Appendix \ref{App-CZDA})
generates a value of $\chi^2=3.6$,
which is larger than that of the asymptotic DA.

The best-fitted linear combination\footnote{%
Dubbed ``diagonal'' in what follows.}
of $a_2,~a_4$, that determines the large axis of the NLO LCSR contour
(see Fig.\ \ref{fig:3-ogur}(a)) and parameterizes its orientation,
is found to be
\begin{equation}
 \label{eq:diameter2}
  a_2+ \frac1{3}\cdot a_4 = 0.143,
\end{equation}
instead of $a_2+ 0.6\cdot a_4= 0.11 \pm 0.03$,
reported in \cite{SchmYa99}.
Note that the SY point also belongs to the diagonal:
$a_2^\text{SY}+ 1/3\cdot a_4^\text{SY} \approx 0.143$.
The coefficient $1/3$ in (\ref{eq:diameter2}) can be predicted
without any fitting; it is solely determined by the structure of the
NLO LCSR (\ref{eq:srggpi}).
Indeed, Eq.\ (\ref{eq:srggpi}) can be rewritten as \cite{SchmYa99}
$$
      A_0(Q^2,\mu^2_\text{SY})+A_2(Q^2,\mu^2_\text{SY})
\cdot a_2(\mu^2_\text{SY})
  +   A_4(Q^2,\mu^2_\text{SY})\cdot a_4(\mu^2_\text{SY})
  =   Q^2 \cdot F^{\gamma^*\gamma\pi}(Q^2)
$$
(to be compared with the fit given in Eq.\ (\ref{eq:diameter2}))
and the discussed coefficient expresses the average value of the ratio
$A_4(Q^2,\mu^2_\text{SY})/A_2(Q^2,\mu^2_\text{SY})$.
Notice that this ratio, averaged over the CLEO data range
$\{Q^2_\text{exp}\}$, amounts to 0.31, while the r.h.s. of
Eq.\ (\ref{eq:diameter2}) is determined by the experimental data on the
form factor.
The coefficient $0.6$, obtained in the SY fit \cite{SchmYa99},
can be associated with the same ratio at the mean value
$\mu_\text{SY}^2 = \va{Q^2_\text{exp}}$.
The ratio $A_4(Q^2)/A_2(Q^2)$ is a concave function in $Q^2$ and,
therefore, its mean value, $\va{A_4(Q^2)/A_2(Q^2)}$, is smaller than
its value,
$ A_4(\mu_\text{SY}^2)/A_2(\mu_\text{SY}^2)$, at the
``mean point'' $\mu_\text{SY}^2$.

\subsection{Sensitivity to input parameters}
As it  turns out, the location of the admissible $a_2,~a_4$ regions
is rather sensitive to the value of the input parameter $\delta^2$.
To illustrate this point, we have repeated the data processing
with an admissible (near its low boundary) value
$\delta^2=0.18~\gev{2}$ (see Appendix \ref{App-delta}).
All contours in Fig.\ \ref{fig:3-ogur}(b)
shift closer to the asymptotic point (\ding{117}),
but their relative positions do not drastically change and,
hence,
the main conclusions (i-iii) of the previous subsection remain valid.
The results of this data processing are presented in
Fig.\ \ref{fig:3-ogur}(b) and in Table \ref{tab-2}.
One appreciates that the hierarchy of the different models (lower
parts of both Tables) with respect to the NLO LCSR best fit does
not change, though the values of $\chi^2$ can change significantly.
Indeed, the point marking the BMS model moves from the
$2\sigma$-deviation level at $\delta^2=0.2~\gev{2}$ (see Table
\ref{tab-1}) inside the $1\sigma$-deviation region near the SY
LCSR point at $\delta^2=0.18~\gev{2}$ (cf. row 5 in Table
\ref{tab-2}).
Therefore, the value of $\delta^2$ and, in general, also the value
of the twist-four term can substantially affect the locations of the
admissible regions.
But all other options, like the CZ DA and the asymptotic DA, remain
excluded at the $2\sigma$--deviation level.

\begin{table}[h]
\caption{Compilation of best fit points shown in Fig.\ \ref{fig:3-ogur}(b)
         with $\delta^2(1~\gev{2})=0.18$~GeV${}^2$.}
\begin{center}
\begin{tabular}{|l|c|c|c|}
\colrule
 ~Best-fit points/models $\vphantom{^|_|}$
         &~$a_2(\mu_\text{SY}^2)$~&~$a_4(\mu_\text{SY}^2)$~&~$\chi^2_\text{(d.o.f.)}$~
 \\ \hline\hline
 ~NLO PT (\ref{eq:NLO}) best fit $\vphantom{^|_|}$
         & $-0.37$             & $+0.34$                   & $0.49$
 \\ \hline
 ~LO LCSR best fit $\vphantom{^|_|}$
         & $+0.006$            & $-0.14$                   & $0.49$
 \\ \hline
 ~NLO LCSR best fit $\vphantom{^|_|}$
         & $+0.17$             & $-0.14$                   & $0.48$
 \\ \hline \hline
 ~SY LCSR~\cite{SchmYa99} $\vphantom{^|_|}$
         & $+0.19$             & $-0.14$                   & $0.51$
\\ \hline
 ~BMS model~\cite{BMS01} $\vphantom{^|_|}$
         & $+0.13$             & $-0.08$                   & $0.56$
  \\ \hline
 ~\text{CZSY DA~\cite{SchmYa99}} $\vphantom{^|_|}$
         & $0.28$              & $-0.009$                  & $2.2$
 \\ \hline
 ~\text{CZ DA~\cite{CZ84}} $\vphantom{^|_|}$
         & $+0.35$             & $-0.006$                  & $4.2$
  \\ \hline 
 ~Asympt. DA  $\vphantom{^|_|}$
         & $-0.006$            & $+0.00$                   & $2.3$
 \\ \hline
\end{tabular}
\label{tab-2}
\end{center}
\end{table}

Let us pause for a moment and turn our attention to a recent paper
by Diehl et al. \cite{DKV01}.
The authors of this work employ a purely perturbative QCD approach
to analyze the CLEO data
without taking into account the twist-four contribution,
i.e., using Eq.\ (\ref{eq:NLO}) with $\delta^2(Q^2)=0$.
They consider this treatment justified given the possible
large uncertainties in estimating the twist-four contribution
(which in the SY procedure is taken to be $\pm 20$\%).
Comparing their results with those of Schmedding and Yakovlev
\cite{SchmYa99}, Diehl et al. correctly note that the relative weights of
$a_2(\mu_0^2)$ and $a_4(\mu_0^2)$
in $F^{\gamma^*\gamma\pi}$ display a much stronger $Q^2$-dependence
than in the leading-twist case with the consequence
that the allowed SY parameter region becomes much smaller
than in their approach.
However, the size of the twist-four contribution is crucial
for accurately extracting the parameters $a_2$ and $a_4$ --
as we have just demonstrated.
Therefore
in our analysis we use a different approach:
the value of $\delta^2$
is connected with the parameter $\lambda_q^2$ of the vacuum non-locality.
We first fix the value of $\lambda_q^2$
and then we allow for the parameter $\delta^2$
to vary in a 10\% range.
The whole uncertainty in $\delta^2$ for the selected range of
$\lambda_q^2=0.4-0.6$~GeV$^2$ amounts then to about 30\%
in accordance with~\cite{Kho01}.
This strategy enables us to use the CLEO data
as a direct measure (a vacuum detector)
to select that model for the QCD vacuum,
which provides the best agreement
between theory and experiment.

\section{Complete two-loop analysis of the CLEO data}
 \label{sec:2loop}
In the previous section we have demonstrated the high sensitivity of
the DA parameters $(a_2, ~a_4)$ to strong radiative corrections
for the form factor, as well as to the scale of the twist-four contribution
(see Fig.\ \ref{fig:3-ogur} a(b) and \cite{BMS01}).
Therefore, to obtain these parameters from the CLEO data in a reliable
way, one should take into account the radiative corrections in the most
accurate possible way.
To this end, we want to improve in this section the accuracy of the
extraction procedure of $(a_2, ~a_4)$ at the NLO level.
A new estimate for $\delta^2$, the magnitude of the twist-four contribution,
is also introduced in the present analysis (see below).
We also briefly discuss an attempt \textsl{to go beyond} the level of
the NLO, having recourse to a recent calculation \cite{MNP01a}
of the radiative correction based on the BLM scale setting.

\subsection{Complete NLO analysis}
 \label{subsec:4.1}
Here we use the complete 2-loop expression for the form factor
$F^{\gamma^*\gamma^* \pi}(Q^2,q^2)$,
given by Eq.\ (\ref{eq:factor2}).
For this reason, we put $\mu^2=Q^2$ in (\ref{eq:SY})
so that for the quantities
\begin{equation}
 \alpha_s(\mu^2)
  \stackrel{\rm RG}{\longrightarrow}
 \alpha_s(Q^2),
 ~~\varphi_\pi(x;\mu^2)
 \stackrel{\rm ERBL}{\longrightarrow}
 \varphi_\pi(x; Q^2)=U(\mu^2 \to Q^2)\varphi_\pi(x; \mu^2),\, \nonumber
\end{equation}
the NLO evolution is implied.
Then, we substitute the spectral density $\rho(Q^2,s;\mu^2=Q^2)$,
derived in \cite{SchmYa99} (see the text below Eq.\ (\ref{eq:SY})),
in LCSR (\ref{eq:srggpi}) to obtain
$F^{\gamma^*\gamma\pi}(Q^2)$ in a regular manner and to fit the CLEO
data over $Q^2\in\{Q^2_\text{exp}\}$.
The evolution
$\varphi_\pi(x; Q^2)=U(\mu^2_\text{SY}
\to Q^2)\varphi_\pi(x; \mu^2_\text{SY})$
is performed \textit{for every point} $Q^2_\text{exp}$, with the aim to
return to the normalization scale $\mu^2_\text{SY}$ and to extract the
DA parameters $(a_2,~a_4)$ at this reference scale.
Stated differently, for every measurement,
$\{Q_\text{exp}^2,F^{\gamma^*\gamma\pi}(Q_\text{exp}^2)\}$,
its own factorization/renormalization scheme has been used so that
the NLO radiative corrections are taken into account in a complete way.

The accuracy of the procedure is, nevertheless,
still limited owing to the mixing of the NLO
and the LO approximations.
Indeed, the value of the twist-four coupling $\delta^2(\mu^2)$, as well
as its RG-evolution with $\mu^2$, are estimated in the LO approximation.
This quantity enters the LCSR formula (\ref{eq:SY}) together with the
NLO-part and can lead to an additional uncertainty.
In order to improve the theoretical accuracy of the values of $a_2$ and
$a_4$, extracted from the CLEO data, one has to re-estimate the twist-four
contribution, $\delta^2(1~\gev{2})$, with a better accuracy.

\begin{figure}[th]
 \centerline{\includegraphics[width=0.47\textwidth]{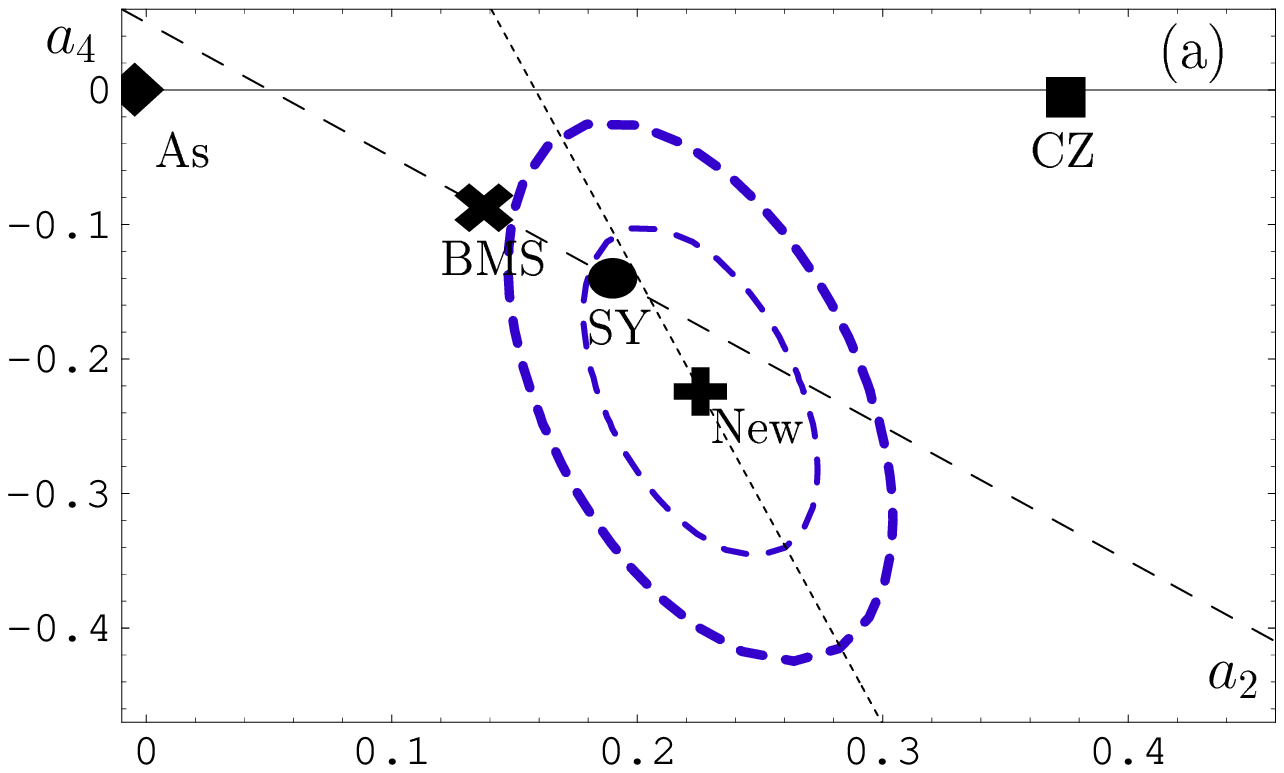}
  \hspace{0.03\textwidth}
   \includegraphics[width=0.47\textwidth]{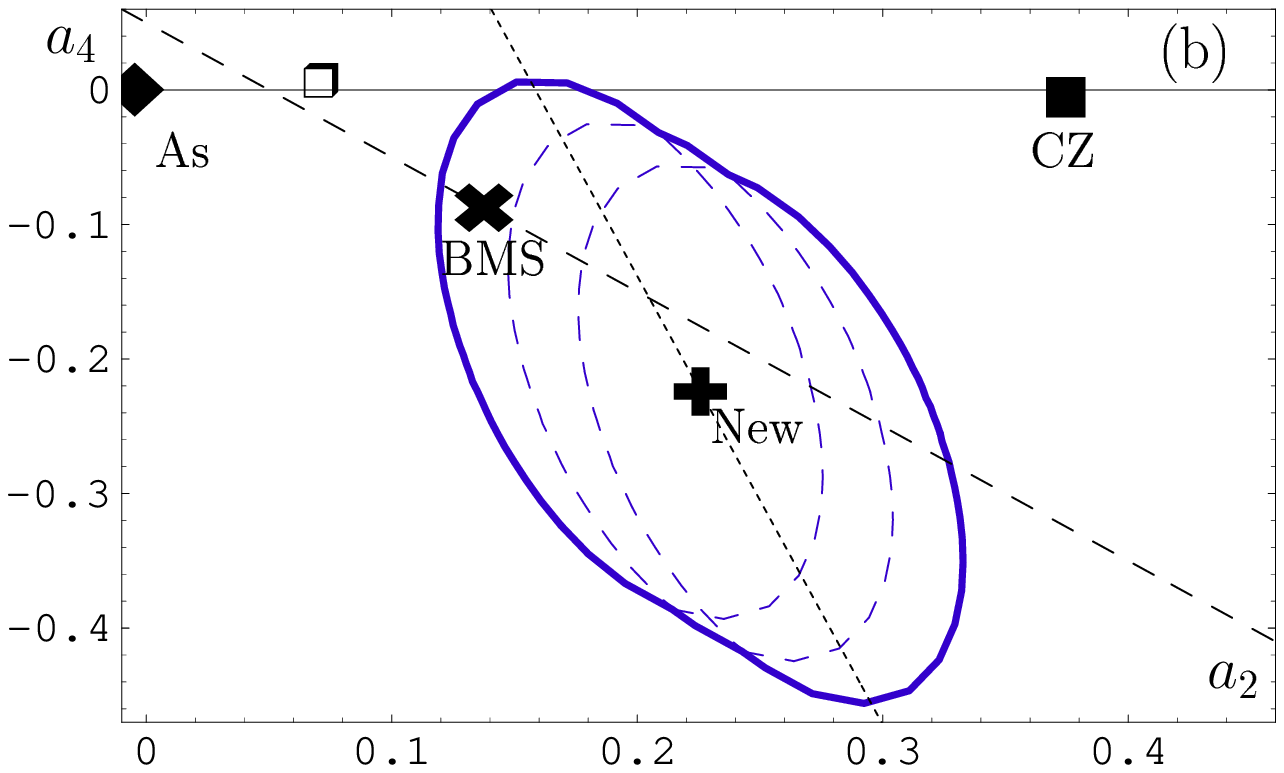}}
    \centerline{\includegraphics[width=0.47\textwidth]{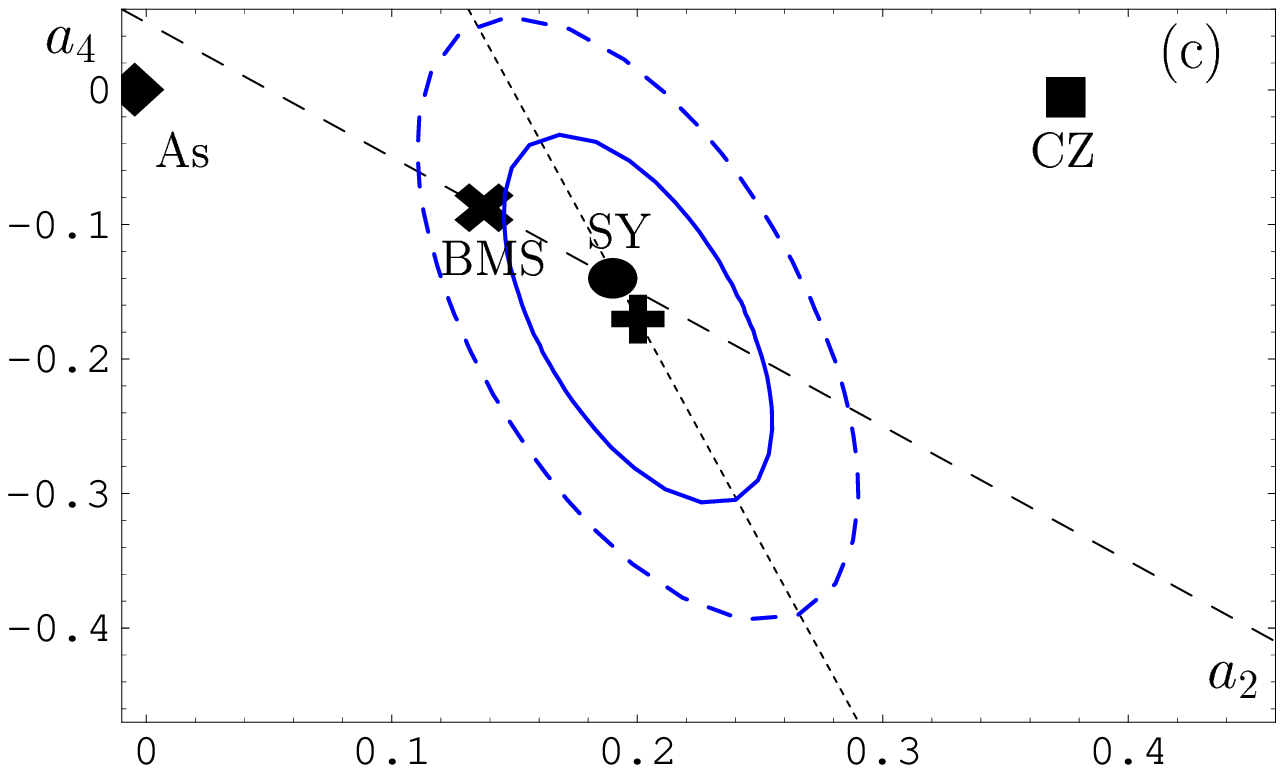}
     \hspace{0.03\textwidth}
      \includegraphics[width=0.47\textwidth]{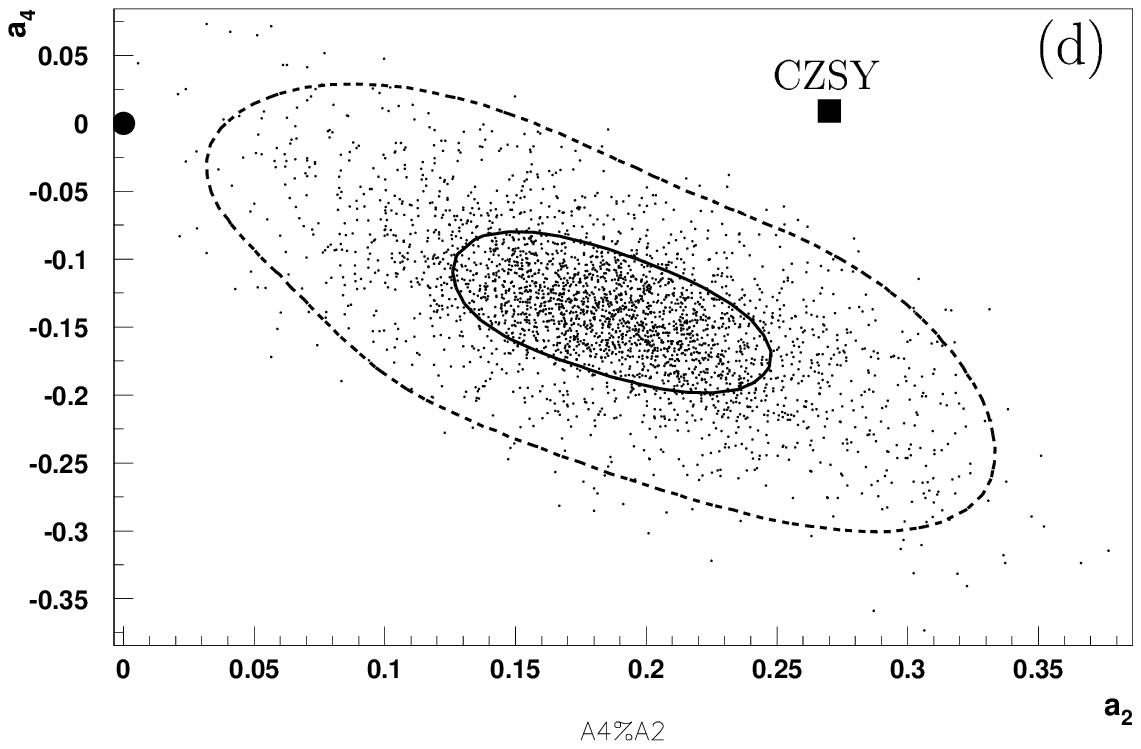}}
       \caption{\label{fig:ogur-new09}
       \footn Results of the analysis of the CLEO data for the form factor
       $F_{\pi\gamma^{*}\gamma}(Q^2)$
       in terms of fiducial areas in the plane ($a_2$,$a_4$).
       (a) We show the best fit point of the new NLO LC SRs (\ding{58},
       termed ``New'', which is associated with the value of the twist-four
       coupling $\delta^2=0.19 ~\gev{2}$, and display by the heavy
       (broken) contour the regions corresponding to a $1\sigma$
       ($2\sigma$)-deviation.
       We also show the SY point (\ding{108}), the BMS point (\ding{54}),
       the asymptotic point (\ding{117}), and the CZ point (\ding{110})
       - all evolved to the SY scale.
       The inclined straight dotted line in the left figure
       is the modified diagonal $a_2 + 0.3 a_4 = 0.159$,
       whereas the inclined straight dashed line is the original diagonal
       $a_2 + a_4 = $\textsc{const}.
       (b) We show three $2\sigma$-regions (dashed contours),
       obtained for different values of the twist-four parameter:
       $\delta^2=0.17,\ 019,\ 0.21~\gev{2}$
       and also their unification: $2\sigma$-contour (thick line).
       All points and straight lines are as in
       Fig.\ \ref{fig:ogur-new09}(a), except \ding{114},
       which denotes a brand-new result
       obtained on a transverse lattice~\protect{\cite{Dal02}}.
       (c) The contours of the previous NLO analysis
       \textit{\`{a} la} SY and the ``old'' best-fit point (\ding{58})
       are presented for comparison
       (cf.\ the lower right corner of Fig.\ \ref{fig:3-ogur}(a)
        and use the scale of Fig.\ \ref{fig:ogur-new09}(a)).
       (d) The original SY plot (note the different scales on the axes
       and the symbol {\tiny \ding{110}} denoting the CZSY DA) is shown
       for comparison.
       All values shown are evaluated at the scale $\mu=2.4$~GeV.
      }
\end{figure}

To summarize, our data processing procedure differs from the SY one in
the following points:
\begin{enumerate}
 \item $\alpha_s(\mu^2)$ is the exact solution of the 2-loop RG
  equation, rather than the approximate (but popular in the HEP community)
  expression (\ref{eq:PDG-coupling}) \cite{PDG2000},
  that was used in the SY analysis.
 \item All logarithms $\ln(Q^2/\mu^2)$ are absorbed into the evolution of
  the pion DA, performed separately at each experimental point
  $Q_\text{exp}$ (compare Eqs.\ (\ref{eq:EvPDA}), (\ref{eq:factor2})
  with Eq.\ (\ref{eq:SY})).
  The corresponding expressions for the form factor are collected in
  Appendix E.
 \item The value of the parameter $\delta^2$ has been re-estimated
  to read $\delta^2(1 \gev{2}) = 0.19\pm0.02~\gev{2}$
  (see Appendix A),
  and this value has been used in the data processing.
\end{enumerate}
This processing of the CLEO data produces the admissible regions,
one of which,
corresponding to $\delta^2=0.19~\gev{2}$,
is shown in Fig.\ \ref{fig:ogur-new09}(a),
where the original SY regions (Fig.\ 6 in \cite{SchmYa99})
are also presented in Fig.\ \ref{fig:ogur-new09}(d)
for the ease of comparison.
To produce the complete $2\sigma$- and  $1\sigma$-contours,
corresponding to  $\delta^2(1 \gev{2}) = 0.19\pm0.02~\gev{2}$,
we need to unite three regions obtained
for different values of the twist-four parameter:
$\delta^2=0.17,\ 019,\ 0.21~\gev{2}$.
This procedure is illustrated in Fig.\ \ref{fig:ogur-new09}(b)
using as an example the $2\sigma$-contour.
Let us remind the reader
in this context that the SY contours are stretched
along the ``LO perturbative'' diagonal $a_2+a_4=$\textsc{const}
(the dashed straight line on the l.h.s. resembles exactly this diagonal)
while the solid (dotted) contours correspond
to the 1~$\sigma$ (2~$\sigma$) regions.
This stretching of the contours appears here
because of the SY manner of the data processing,
namely, because the theoretical uncertainties of the
input parameters were also involved in the statistical analysis.
\begin{table}[h]
\caption{Compilation of best-fit points shown in Fig.\ \ref{fig:ogur-new09}(a)
         with $\delta^2(1~\gev{2})=0.19$~GeV${}^2$.}
\vspace{3mm}
\begin{center}
\begin{tabular}{|l|c|c|c|c|}
\colrule
 ~Best-fit point/models $\vphantom{^|_|}$
  &~$a_2(\mu_\text{SY}^2)$~&~$a_4(\mu_\text{SY}^2)$~&~$\chi^2_\text{(d.o.f.)}$~
                                                             &~$a_2+ 0.3\cdot a_4$~
 \\ \hline\hline
 ~New  NLO LCSR best fit $\vphantom{^|_|}$
  & $+0.23$                & $-0.22$                & $0.47$ & $0.159$
 \\ \hline \hline
 ~SY NLO LCSR~\cite{SchmYa99} $\vphantom{^|_|}$
  & $+0.19$                & $-0.14$                & $0.59$ & $0.153$
 \\ \hline
 ~BMS model~\cite{BMS01} $\vphantom{^|_|}$
  & $+0.14$                 & $-0.09$               & $1.0$ & $0.113$
 \\ \hline
 ~Asymptotic model $\vphantom{^|_|}$
    & $-0.002$             & $+0.00$                & $4.3$  & $-0.00$
 \\ \hline
 ~CZ model~\cite{CZ84} $\vphantom{^|_|}$
    & $+0.43$              & $-0.003$               & $4.3$  & $0.43$
 \\ \hline
\end{tabular}
 \label{tab-3}
\end{center}
\end{table}
The new best-fit point (\ding{58}, ``New''), as well as the whole
$\sigma$-contours themselves appear to be displaced in
Fig.\ \ref{fig:ogur-new09}(a) (approximately) along the new diagonal
(cf.\ Eq.\ (\ref{eq:diameter2})),
\begin{equation}
 \label{eq:diameter-new}
   \quad a_2 + 0.3\cdot a_4 = 0.16 \pm 0.007\, .
\end{equation}
In Fig.\ \ref{fig:ogur-new09}(c) we present for comparison the contours
of the previous NLO analysis in the sense of SY (low right corner of
Fig.\ \ref{fig:3-ogur}(a)) drawn, however, at the scale of
Fig.\ \ref{fig:ogur-new09}(a).
The positions of the best-fit points and models are provided in
Table \ref{tab-3}.

It should be clear from our discussion that these new contours
are somewhat smaller than the previous ones
(Fig.\ \ref{fig:ogur-new09}(c)), but slightly larger than the original
SY ones (Fig.\ \ref{fig:ogur-new09}(d)), and show another orientation
along the diagonal Eq.\ (\ref{eq:diameter-new}).
The difference between the new regions,
determined in the present analysis,
and those of the SY one is remarkable.
For instance, the SY  point appears now near the boundary and inside
the $1\sigma$-region in Fig.\ \ref{fig:ogur-new09}(a).
Moreover, the preliminary (i.e., for $a_4=0$) SY  best-fit point,
$a'_2=0.12\pm 0.03$ \cite{SchmYa99}, and the phenomenological estimates
for ($a_2,~a_4$), presented in \cite{BKM00},
($a_4=0,~a_2~(1~\gev{2})=0.1 \pm 0.1$),
lie on the boundary of the united $2\sigma$-region
(see Fig.\ \ref{fig:ogur-new09}(b)).

\subsection{Beyond the NLO approximation: effects from BLM scale-setting}
\label{subsec:4.2}
The renormalization scale $\mu_\text{R}^2$ in Eq.\
(\ref{eq:factor2}) can be fixed by a NNLO calculation of $T_2$
following the BLM prescription.
Recently, the NNLO contribution to $C_2$, proportional to $b_0$ and
required for the BLM scale setting, was obtained in \cite{MNP01a} for a
kinematics with $q_2^2=0$.
As an exercise, let us perform the new fit to obtain the scale
$\mu_\text{R}^2=\mu^2_\text{BLM}$ and the best-fit point
$\{a^\text{BLM}_2, a^\text{BLM}_4 \}$ for the NLO expression given
by Eq.\ (\ref{eq:NLO}).
We follow the same procedure as in Sec.\ \ref{subsec:3.1}, replacing this
time $\as(Q^2) \to
\as(\mu^2_\text{BLM}(a_2,a_4,Q^2))$ in Eq.\ (\ref{eq:NLO}), where
for $\mu^2_\text{BLM}(a_2, a_4,Q^2)$ Eq.\ (7.7) from \cite{MNP01a}
is used.
As it turns out, practically for  all points in the considered domain in
the lower half-plane $a_4 \leq 0$, the BLM setting leads to the condition
$\mu^2_\text{BLM} \ll Q^2$, in conformance with the results of \cite{MNP01a}.
Therefore, for this region, the BLM setting seems to rule out predictions
from the NLO perturbation theory.

Only for points within a rather thin strip in the upper half-plane $a_4 > 0$
(cf. Eqs.\ (7.2a), (7.7) in \cite{MNP01a}), the BLM setting gives
$\mu^2_\text{BLM} > Q^2$.
Interestingly,
the case discussed in Sec.\ \ref{subsec:3.1} (ii)
and based on Eq.\ (\ref{eq:NLO})
belongs just to this thin strip.
The corresponding $(a_2,a_4)$ values are displayed below for
comparison together with the initial result (second row) without
the BLM setting.
\begin{eqnarray} \label{blmNLO}
a^\text{BLM}_2(\mu_\text{SY}^2) = -0.55, &\ \
  a^\text{BLM}_4(\mu_\text{SY}^2) = 0.48 &
  \mu^2_{R}=
  \mu^2_\text{BLM} \approx Q^2/0.35  \\
a_2(\mu_\text{SY}^2) = -0.44, &\ \
 ~~~~~a_4(\mu_\text{SY}^2) = 0.40 \ &
  \mu^2_{R} = Q^2\, . \nonumber
\end{eqnarray}
To calculate the imaginary part
$\mathbf{Im}
  F_\text{QCD}^{\gamma^*\gamma^*\pi}(Q^2,s)$,
used in Eq.\ (\ref{eq:srggpi}), one needs to know the
NNLO contribution proportional to $b_0$ for $q_2^2\neq0$, which is
still not computed.
For this reason, the results obtained with the BLM scale setting
fall out of the region of the NLO LCSR analysis.
The calculation of the complete NNLO contribution or, at least, its
convolution with $\varphi_{as}$ is a very demanding task that has
not been accomplished yet.

\section{Pion DA from QCD SR vs CLEO data}
\label{sec:pionDA}
\begin{figure}[h]
 \centerline{\includegraphics[width=0.33\textwidth]{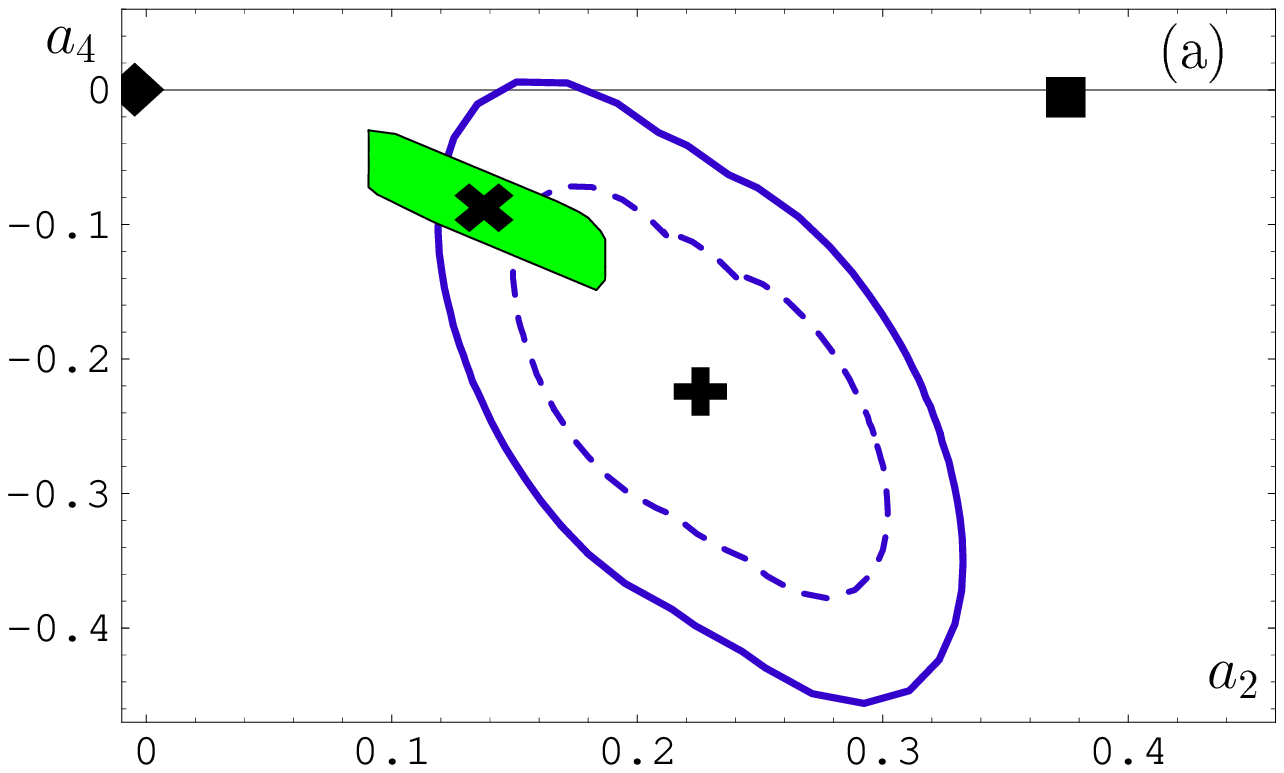}%
  ~\includegraphics[width=0.33\textwidth]{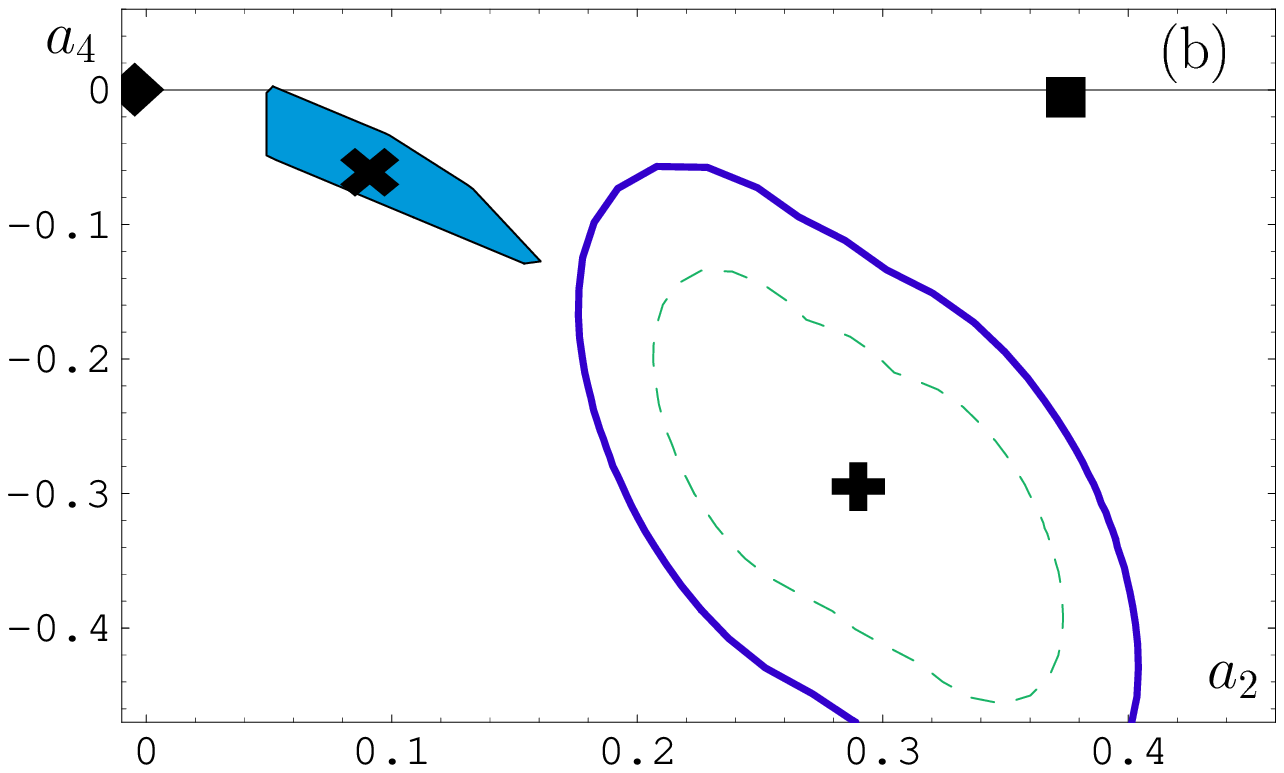}%
  ~\includegraphics[width=0.33\textwidth]{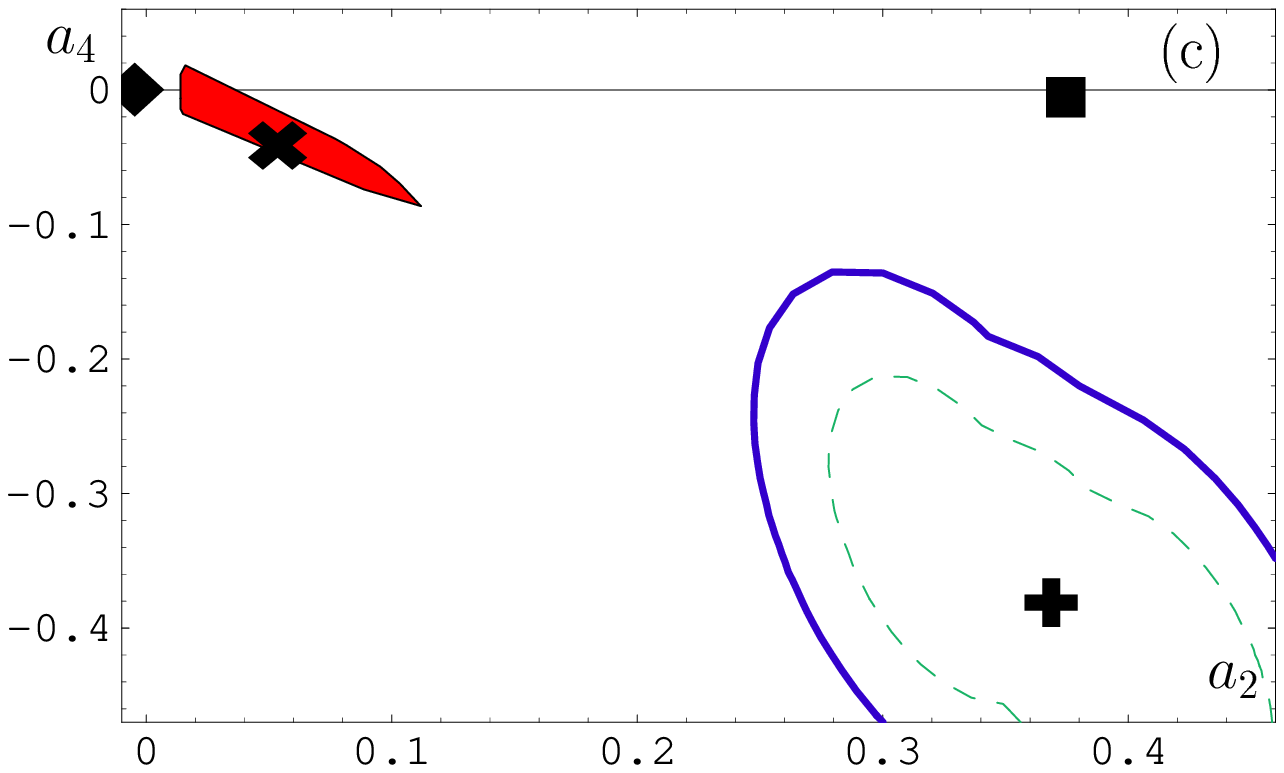}}%
  \caption{\label{fig:3NColor}
   \footn
   Three $2\sigma$-contours of the admissible regions
   following from the analysis of the CLEO data for different values
   of $\delta^2$:
   (a) -- for $\lambda^2_q=0.4~\gev{2}$ and $\delta^2=(0.19\pm0.02)~\gev{2}$;
   (b) -- for $\lambda^2_q=0.5~\gev{2}$ and $\delta^2=(0.235\pm0.025)~\gev{2}$;
   (c) -- for $\lambda^2_q=0.6~\gev{2}$ and $\delta^2=(0.29\pm0.03)~\gev{2}$.
   Solid lines in all figures enclose the $2\sigma$-contours,
   whereas the $1\sigma$-contours are enclosed by dashed lines.
   The three slanted and shaded rectangles represent the constraints
   on ($a_2,~a_4$) posed by the QCD SRs \protect\cite{BMS01}
   for corresponding values of $\lambda^2_q=0.4,~0.5,~0.6$~\gev{2}
   (from left to right).
   All values are evaluated at $\mu=2.4$~GeV after evolution.
   The marked points are explained in Fig.\ \ref{fig:ogur-new09},
   except for the point marked with \ding{54}.
   This point represents here the BMS solution,
   which corresponds to a particular value of $\lambda_q^2$.}
\end{figure}

Let us now turn to the important topic of whether the CLEO data is
consistent with the non-local QCD SR results for $\varphi_{\pi}$.
We present in Fig.\ \ref{fig:3NColor} the results of the data analysis
for three central values of the coupling
$\delta^2(\mu^2=1~\gev{2})=0.19, 0.235, 0.29~\gev{2}$,
which in turn correspond to three admissible values
of the vacuum non-locality parameter
$\lambda_q^2=0.4, 0.5, 0.6~\gev{2}$.
For each value of $\lambda_q^2$ from this ensemble,
we define the corresponding central value of $\delta^2$
and its uncertainty (for details, see Appendix~\ref{App-delta}).
Then, we process the CLEO data as described in the previous section
and obtain
the complete $2\sigma$- and $1\sigma$-contours
on the plane $(a_2,a_4)$,
following from the CLEO experiment.
An example of these regions is represented
in Fig.\ \ref{fig:ogur-new09}(b) and
is also displayed in Fig.\ \ref{fig:3NColor}(a),
where the $2\sigma$-contour is shown as a solid line
and the $1\sigma$-contour as a dashed one.

The task now is to compare these new constraints
with those following
from the QCD SRs with nonlocal condensates.
We have established in \cite{BMS01}
that a two-parameter model $\varphi_{\pi}(x;a_2, a_4)$
really enables one
to fit all the moment constraints for
$\langle \xi^{N} \rangle_\pi$
that result from NLC QCD SRs (see \cite{MR89} for more details).
The only parameter entering the NLC SRs is the correlation scale
$\lambda^2_q$ in the QCD vacuum, known from nonperturbative
calculations and lattice simulations (for a discussion and references,
see Appendix \ref{App-delta}).\\

The three slanted and shaded rectangles in Fig.\ \ref{fig:3NColor}
are the constraints on the Gegenbauer coefficients ($a_2,~a_4$)
resulting from the NLC QCD SRs at different values of
$\lambda^2_q=0.4, ~0.5, ~0.6$~\gev{2}, \cite{BMS01,BM02}.
The overlap of the displayed regions in Fig.\ \ref{fig:3NColor}
can serve as a means of determining
the appropriate value of $\lambda^2_q$.
In fact, one may conclude that the value $\lambda^2_q=0.4 $~\gev{2}
is more preferable relative to the higher values of $\lambda^2_q$.
It should be noted, however, that even for this lowest value of
that scale, the agreement with the constraints in
Fig.\ \ref{fig:3NColor} is rather moderate
and of similar quality
as using the SY constraints \cite{BMS01,BM02}.
It is tempting to test even smaller values of $\lambda^2_q$
than $0.4$~\gev{2} in the NLC SR
as an attempt to improve the agreement
with the CLEO constraints in Fig.\ \ref{fig:3NColor}.
But such values appear to be at the lower limits for  the
$\lambda^2_q$ estimates from non-perturbative approaches
(see Appendix A).
Furthermore, the NLC SR becomes unstable at such low values of
$\lambda^2_q$.
As a result, the accuracy of the DA moments is rather poor and
the final constraint on ($a_2,~a_4$) becomes unreliable.
Taking into account all these arguments, we think that an
improvement of the ingredients of the NLC ansatz may
provide a better agreement with the CLEO data
than just using
$\lambda^2_q< 0.4$~\gev{2}.

\section{Conclusions}
\label{sec:concl}
In this paper we have studied the theoretical predictions for
the pion transition form factor
$F^{\gamma^*\gamma\pi}(Q^2)$
in comparison with the CLEO experimental data \cite{CLEO98}
on this form factor.
We have presented a full analysis of this data and contrasted the
results with those found in the context of QCD LCSR at the NLO level.
In this way, we have revised and improved the procedure of analyzing
the CLEO data, first performed by Schmedding and Yakovlev in
\cite{SchmYa99}.
The main goal has been to obtain constraints on the shape of the pion
DA of twist-2, $\varphi_{\pi}(x;a_2,a_4)$
in the most accurate way.
The values of the crucial parameters, viz. the twist-four coupling
$\delta^2(\mu^2)$ and $\as(\mu^2)$, involved in this procedure,
have been treated more accurately than in previous approaches.
The main findings may be summarized as follows.
\begin{enumerate}
\item We have tested different kinds of approximations to calculate
$ F^{\gamma^*\gamma\pi}(Q^2)$
and revealed how the size of $\as$ and the twist-four corrections
can affect the admissible regions of the DA parameters $(a_2, a_4)$.

\item New admissible regions for the $(a_2, a_4)$ parameters,
 see Figs.\ \ref{fig:ogur-new09}(b), \ref{fig:3NColor}(a),
 -- different from those in \cite{SchmYa99} --
 have been obtained, though the constraints do not change drastically
 in the sense that the initial SY best-fit point still belongs
 to a $1\sigma$-deviation region (CL=68\%) in this space,
 whereas the  CZ DA and also the asymptotic one
 are definitely excluded at the level of a $2\sigma$-deviation criterion (CL=95\%).
 Moreover, one may appreciate by comparing Figs.\ \ref{fig:ogur-new09}(a, b)
 with Fig.\ \ref{fig:ogur-new09}(d)
 that this exclusion
 with respect to the asymptotic DA
 becomes even  more pronounced using our data processing.

\item The bunch of admissible pion DAs,
 $\varphi^{BMS}_{\pi}(x;a_2,a_4)$, corresponding to the estimate
 $\lambda^2_q=0.4$ GeV$^{2}$, that was constructed within the framework
 of QCD SRs with nonlocal condensates in \cite{BMS01},
 compares well (at the $2\sigma$--level)
 with the new more restrictive constraints
 obtained in the present investigation
 as Fig.\ \ref{fig:3NColor} demonstrates.
 In addition, half of the calculated admissible region intersects with
 the $1\sigma$ domain as well, see Fig.\ \ref{fig:3NColor}(a).
\end{enumerate}
\vspace{0.2cm}

\begin{acknowledgments}
This work was supported in
part by the Russian Foundation for Fundamental Research
(contract 00-02-16696), INTAS-CALL 2000 N 587,
the Heisenberg--Landau Programme (grant 2002-15),
and the COSY Forschungsprojekt J\"ulich/Bochum.
We are grateful to A.\ Kotikov, A.\ Nagaitsev, K.\ Passek,  A.\ Radyushkin,
D.\ V.\ Shirkov, A.\ Sidorov, M.\ Strikman, and O.\ Teryaev
for discussions and O.\ Yakovlev for correspondence.
One of us (A.~B.) is indebted to Prof.\ Klaus Goeke
for the warm hospitality at Bochum University,
where this work was partially carried out.
\end{acknowledgments}

\newpage
\begin{appendix}
\appendix

\section{Revision of the QCD SR results for $\delta^2$}
 \renewcommand{\theequation}{\thesection.\arabic{equation}}
  \label{App-delta}\setcounter{equation}{0}
The coupling $\delta^2(\mu^2)$ was originally estimated in \cite{NSVZ84}
and found to be $\delta^2(\mu^2=1~\gev{2}) = 0.2 \pm 0.02~\gev{2}$.
Here, we re-analyze the QCD SR for $\delta^2$,
derived in \cite{OPiv88},
which is based on a non-diagonal correlator
of the quark-gluon and quark (pseudoscalar) currents.
This SR relates $\delta^2$ to $\lambda_{q}^{2}$ and determines the value
of the ratio
$\Ds r = \lambda_{q}^{2}/(2\delta^2)$.
Evaluating the SR leads to the estimate $r > 1$ and consequently to
$\lambda_{q}^{2}/2 >\delta^2$.
Moreover, $r$ is rather sensitive to the size of the radiative
corrections.
In this work, we use $\Lambda_{3}^\text{LO}=312~\Mev$, obtained
recently in a DIS fit of the CCFR data in \cite{KPS02} that leads to
$\as^\text{LO}(1~\gev{2})\approx 0.59$.
The same sort of analysis in the NLO approximation leads to the estimate
$\Lambda_{3}^\text{NLO} = 410~\Mev$\footnote{%
Using the values
$\Lambda_{4}^\text{LO}=265~\Mev$ and $\Lambda_{4}^\text{NLO} = 340~\Mev$
given in \cite{KPS02} for $N_f=4$, we re-calculated the values for
$N_f=3$, i.e.,
$\Lambda_{3}^\text{LO}=312~\Mev$ and $\Lambda_{3}^\text{NLO} = 410~\Mev$.}
that is indeed not far from the standard value $380~\Mev$
(Appendix \ref{App-C2}).

To determine $\delta^2$, we first fix the parameter $\lambda_{q}^{2}$
by employing the ``conservative estimate''
$\lambda_{q}^{2}(\mu^2=1~\gev{2})=0.4$ GeV$^{2}$.
In QCD the value of this parameter
was estimated in the QCD SR approach~\cite{BI82}
and also using lattice data~\cite{BM02}:
\begin{equation}
\label{lambda}
 \lambda_{q}^{2}= \frac{\va{\bar{q}(0)\nabla^{2} q(0)}}
                   {\va{\bar{q}(0)q(0)}}
            \begin{array}{c}
             \text{\footn in chiral}\\
             =\\
             \text{\footn limit}
             \end{array}
              \frac{\va{\bar{q}(0)
                        \left(i g\,\sigma_{\mu \nu}G^{\mu \nu}
                        \right)q(0)}}
                   {2\va{\bar{q}(0)q(0)}}
   = \left\{\begin{array}{ll}
         0.4 \pm 0.1~\gev{2}& \text{\cite{BI82,OPiv88}}\\
         0.4 - 0.5~\gev{2}&\text{\cite{BM02}}
             \end{array}
     \right.\ .
\end{equation}
A brief review of the different estimates of $\lambda_{q}^{2}$
is given in \cite{BM02}.

The evaluation of the SR for the quantity $ r$,
\begin{equation} \label{eq:r}
\Ds
  \frac{\lambda_{q}^{2}}{2\delta^2}\equiv  r(M^2, \varrho, s_0)
  = \frac{\Ds 1 - \varrho \cdot \exp\left(-m^2_{\pi'}/M^2\right)}
         {\Ds 1 +  \frac{2\pi^2}{9}
                   \frac{\va{G^2}}{\lambda_{q}^{2} M^2}
                -  \frac{2\as(1~\gev{2})}{3\pi}
                   \frac{M^2}{\lambda_{q}^{2}}
                   \left[1 - \exp(-s_0/M^2)\right]}
\end{equation}
\begin{figure}[h]
 \noindent
   \centerline{\includegraphics[width=0.6\textwidth]{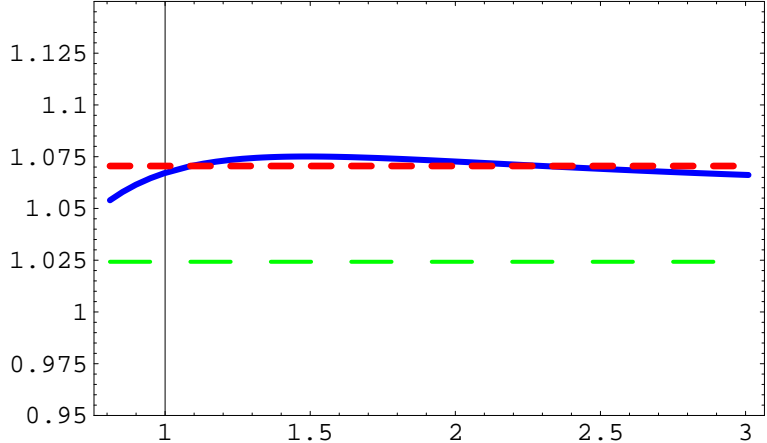}}
    \caption{\label{fig:delta}\footn
    The solid line denotes $r(M^2, \varrho,s)$ for the best-fitted
    parameters  $\varrho=1.1,~s_0=3.24$.
    The short-dashed line corresponds to the value $r=1.07$,
    whereas the long-dashed line corresponds to $r$
    calculated with $\Lambda_{3} = 200$~MeV.}
\end{figure}
for the standard value of the gluon condensate
$\va{G^2}\equiv \va{\as GG}/\pi=0.012~\gev{4}$, \cite{SVZ},
and with the fitting parameters,
i.e., the coupling to the $\pi'$, $\varrho \approx 1.1$,
and the duality interval $s_0=3.24~\gev{2}$ yields
\begin{equation}  \label{eq:Lam312}
\ r=1.07 (\pm 0.01) \Rightarrow \delta^2 \approx 0.19~\gev{2}~
 \mbox{for }~ \Lambda_{3}^\text{LO}=0.312~\Mev \, .
\end{equation}

The stability of the SR (\ref{eq:r}) with respect to the Borel
parameter $M^2$ is rather good, as the solid line in
Fig.\ \ref{fig:delta} clearly effects.

Note that adopting the popular option
$\Lambda_{3}^\text{LOpop} = 200$~MeV,
it simply imitates the $\as^\text{NLO}(\mu^2)$ behavior at
intermediate scales $\mu^2$, providing $ r=1.02 (\pm 0.01)$ and
$\delta^2 \approx 0.20~\gev{2}$.
This old estimate \cite{OPiv88} is also presented here for comparison
(cf.\ long-dashed line in Fig.\ \ref{fig:delta}).
The parameter $r$ is also rather sensitive to the value of $\va{G^2}$.
For example, the new estimate of the mean value
$\va{G^2}=0.009~\gev{4}$, suggested quite recently in \cite{Ioffe02},
leads to $r\simeq 1.1$ and $\delta^2 \simeq 0.18~\gev{2}$.
Taking into account the last estimate, we derive the uncertainty
$\Delta_G\delta^2 \simeq 0.01~\gev{2}$.
The other uncertainties, inherent in the SR method (\ref{eq:r}),
result to an overall effect of the same order: viz.,
$\Delta \delta^2 \simeq 0.01~\gev{2}$.
Finally, we can establish for $\delta^2$ an accuracy:
$\delta^2=(0.19 \pm 0.02)~\gev{2}$
for $\lambda_q^2=0.4~\gev{2}$.

In the same way we obtain from~(\ref{lambda}) the corresponding central values
and uncertainties of $\delta^2$
for the higher value of $\lambda_q^2=0.5~\gev{2}$
($\delta^2=0.235\pm0.025$)
and analogously for the trial value $\lambda_q^2=0.6~\gev{2}$
($\delta^2=0.29\pm0.03~\gev{2}$),
the latter being of interest due to instanton models
~\cite{DEM97,Pra01}.

The one-loop anomalous dimension of $\delta^2$ is
$\gamma_{T4} = 32/9$ (see, for instance, \cite{BKM00}).
On the other hand, the one-loop scale dependence of
$\delta^2(\mu^2)$ is given by
\begin{equation}\label{eq:Ev1L}
 \delta^2(\mu^2)
   = \left[\frac{\as(\mu^2)}{\as(\mu^2_0)}
     \right]^{\gamma_{T4}/b_0} \delta^2(\mu^2_0)\, .
\end{equation}

\section{CZ DA normalization point}
 \renewcommand{\theequation}{\thesection.\arabic{equation}}
  \label{App-CZDA}\setcounter{equation}{0}
The authors of \cite{SchmYa99} used as a normalization point for
the CZ DA the scale $\mu_1^2=0.5~\gev{2}$,
which is significantly larger
than the original one used by Chernyak and Zhitnitsky:
$\mu_\text{CZ}^2=0.25~\gev{2}$. This latter and rather low
normalization point is due to the fact
that CZ employed the description of charmonium decays,
where the characteristic virtuality of the pion
is indeed of this low order as $\mu^2_\text{CZ}$.
Furthermore, in order
to construct their model at such a low scale,
they evolved the 2nd moments, determined at a
scale of 1.5~GeV$^2$, down to this scale using 1-loop evolution
equations with $\Lambda_\text{QCD}=100~\Mev$. The result is the
well-known CZ DA:
\begin{equation}
 \label{eq:CZ_DA}
  \varphi_\text{CZ}(x;\mu_\text{CZ}^2)
  = 6 x (1-x)
    \left[1 + \frac{2}{3}C^{3/2}_2(2x-1)
    \right]
  = 30 x (1-x) (1 - 2 x)^2\, .
\end{equation}
However, if one wants to know the shape of this model at another
scale, one has to evolve it to that scale.
But a natural question arises:
what evolution equation should be used to do that?

From our point of view, the best solution would be to determine
once and for all the value of the second Gegenbauer coefficient of
the CZ DA at the standard QCD SR scale, $\mu_0^2=1~\gev{2}$,
which is, also according to the CZ arguments, rather close to the
scale of the second moment, 1.5~GeV$^2$.
In this sort of determination, one needs to evolve from the CZ
scale $\mu_\text{CZ}^2$ to the scale $M^2=1.5~\gev{2}$ of the
second moment $\va{\xi^{2}}_\text{CZ}$ using the same 1-loop
evolution equations (with $\Lambda_\text{QCD}=100~\Mev$) as in
their original paper \cite{CZ82}.
This produces
\begin{equation}
 \label{eq:CZ_a2(1.5)}
  a_2^\text{CZ}(\mu^2=1.5~\gev{2}) = 0.51\,.
\end{equation}
But after restoring this way the CZ model
at a scale of 1.5~GeV$^2$,
we should use for the evolution to 1~GeV$^2$
the actual value of the 1-loop QCD scale,
i.e., 312~MeV.
This gives
\begin{equation}
 \label{eq:CZ_a2}
  a_2^\text{CZ}(\mu^2=1~\gev{2}) = 0.56\, .
\end{equation}

\section{Radiative corrections }
 \renewcommand{\theequation}{\thesection.\arabic{equation}}
  \label{App-RC}\setcounter{equation}{0}
\subsection{Structure of the NLO amplitude $\mathbf{T_1}(Q^2,q^2;\mu^2;x)$ }
Here we present diagram per diagram the results of the NLO $T_1$ calculation,
performed in \cite{KMR86} in the Feynman gauge
using the ``naive-$\gamma_5$ scheme'' \cite{DaCh81, MNP01a}.
\addtocounter{footnote}{1}  
\begin{table}[h]
\caption{NLO results for individual diagrams$^6$}
 \label{tab-5}
  \begin{tabular}{p{\tabcolf}|p{\tabcols}|p{\tabcolt}}\hline
   &  \multicolumn{2}{c}{$\vphantom{^{\Big|}_{\Big|}}%
     T_1 = N_{T}\big[\log$-part\ $+\ C_1\big]$}
   \\ \cline{2-3}
   &
   &
   \\
     $$\mbox{Diagram}$$ & $$\ln\Big[\bar{Q}^2/\mu^2 \Big]\mbox{-part}$$
   & $$C_1(x,\omega)$$
   \\ \hline
    \begin{minipage}{0.95\tabcolf}
     $$\includegraphics[width=\textwidth]{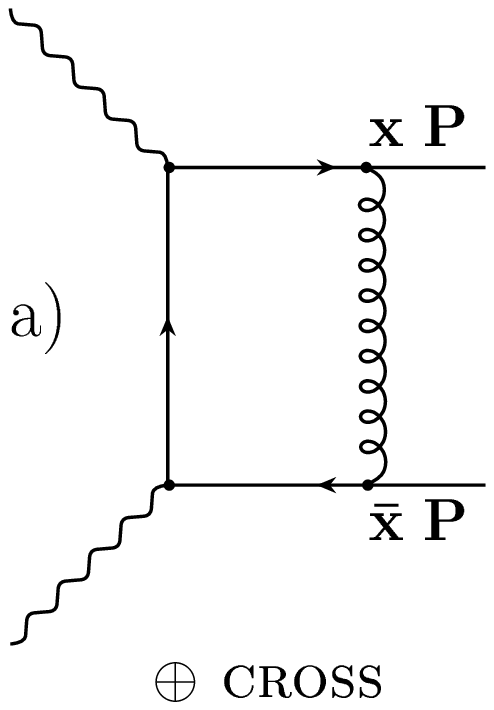}$$
    \end{minipage}
   &~
    \begin{minipage}{0.97\tabcols}
     \begin{eqnarray}
      \ln\Big[\frac{\bar{Q}^2}{\mu_{F}^2}\Big]
       C_0(u,\omega)\convo{u}V_a(u,x)_{+}
      \nonumber\\
      +\ \ln\Big[\frac{\bar{Q}^2}{\mu_{F}^2}\Big] C_0(x,\omega)
      ~~~~~~~~~~~~\nonumber
     \end{eqnarray}
    \end{minipage}
   &
    \begin{minipage}{0.95\tabcolt}
     $$ C'_0(u,\omega)\convo{u}V_a(u,x)_{+}\
     +\ C'_0(x,\omega) $$
    \end{minipage}
   \\ \hline
    \begin{minipage}{0.95\tabcolf}
     $$\includegraphics[width=\textwidth]{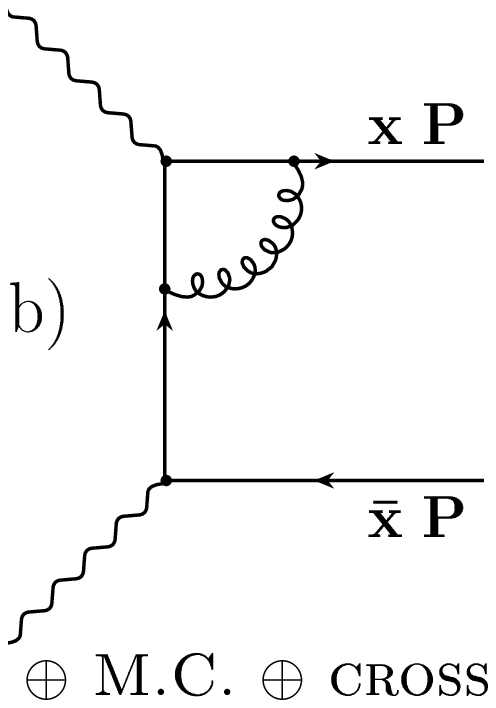}$$
    \end{minipage}
   &~~
    \begin{minipage}{0.95\tabcols}
     \begin{eqnarray}
       \ln\Big[\frac{\bar{Q}^2}{\mu_{F}^2}\Big]
        C_0(u,\omega)\convo{u} V_b(u,x)_{+}
      \nonumber \\
      -\ 2\ln\Big[\frac{\bar{Q}^2}{\mu_{R}^2}\Big] C_0(x,\omega)
      ~~~~~~~~~~
      \nonumber
     \end{eqnarray}
    \end{minipage}
   &\hspace*{1.4mm}\begin{minipage}{0.98\tabcolt}
     \begin{eqnarray}
      \Big[C'_0(u,\omega)-C_0(u,\omega)\Big]\convo{u}V_b(u,x)_{+}
      -\ 2C_0(x,\omega)
      \nonumber\\
      +\ \Big\{\ln[1+\omega(\bar{u}-u)]
               \convo{u}
               \frac{V_b(u,x)_{+} + K(u,x)}
                    {\bar{Q}^2[1+\omega(\bar{x}-x)]}
      \nonumber\\
      +\ (x \to \bar{x}) \Big\}~~~~~~~~~~~~~~~~~~~~~~~~~~~~~~~~~~~
      \nonumber
     \end{eqnarray}
    \end{minipage}
   \\ \hline    
   \begin{minipage}{0.95\tabcolf}
     $$\includegraphics[width=\textwidth]{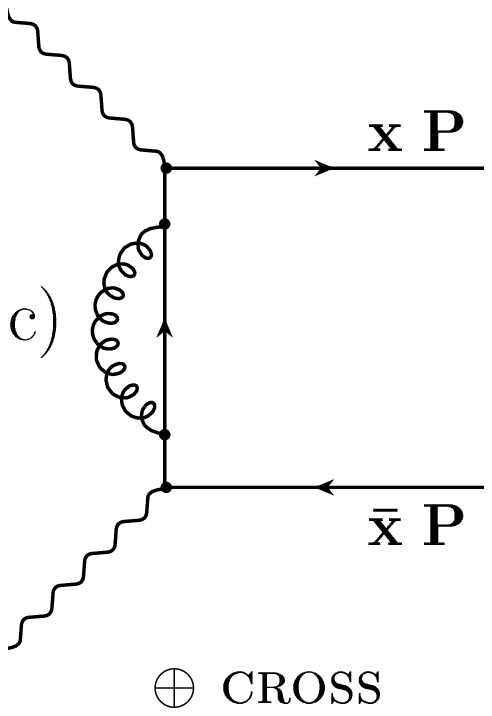}$$
    \end{minipage}
   &
    \begin{minipage}{0.95\tabcols}
     $$\ln\Big[\frac{\bar{Q}^2}{\mu_{R}^2}\Big] C_0(x,\omega)$$
    \end{minipage}
   &
    \begin{minipage}{0.95\tabcolt}
     $$ C'_0(x,\omega)-C_0(x,\omega)$$
    \end{minipage}
   \\ \hline
  \end{tabular}
\end{table} 
\footnotetext{M.C. near a diagram denotes the mirror conjugated diagram,
 cross - the diagram with crossed quark legs.}
To make the presentation more compact, the average virtuality $\bar{Q}^2$
and the asymmetry parameter $\omega$
\begin{equation}
 \bar{Q}^2 = \frac{Q^2+q^2}{2}\,,~~~
 \omega = \frac{Q^2-q^2}{Q^2+q^2}\,,
\end{equation}
have been used, employing also the notation $\bar{x}\equiv 1-x$.

The results are expressed in terms of the LO coefficient function
$C_0(x,\omega)$ (see Eq.\ (\ref{eq:factor0})) and its logarithmic
modification $C'_0(x,\omega)$, naturally appearing in NLO calculations,
\begin{equation}
C_0(x,\omega) =
  \frac{1}{\bar{Q}^2}
   \frac{1}{1+ \omega(\bar{x}-x)}
 + (x\to\bar{x})\,;~~~
C'_0(x,\omega) =
  \frac{1}{\bar{Q}^2}
   \frac{\ln[1+\omega(\bar{x}-x)]}{1+ \omega(\bar{x}-x)}
 + (x\to\bar{x}),
\end{equation}
and their convolutions with $V_a$ and $V_b$, the latter being parts
of the LO ERBL kernel,
\begin{eqnarray}
  V_a(y,x) =
  2C_\text{F}\,\theta(x-y)\frac{y}{x}
   + (x\to\bar{x}, y\to\bar{y})\,;~~
  V_b(y,x) =
  2C_\text{F}\,\frac{y}{x}\frac{\theta(x-y)}{x-y}
   + (x\to\bar{x}, y\to\bar{y})\,;\nonumber
\end{eqnarray}
\begin{equation}
 V_0(y,x)
  =  V_a(y,x)_{+} + V_b(y,x)_{+}
  =  2C_\text{F}
      \left[\theta(x-y)\frac{y}{x}
             \left(1 + \frac{1}{x-y}\right)
           + (x\to\bar{x}, y\to\bar{y})
      \right]_{+}\,.
\label{eq:AppB_kernel-0}
\end{equation}
\noindent Here, the $_{+}$-form of a distribution $V(y,x)$ is defined in the common way:
\begin{equation}
 V(y,x)_{+}
 \equiv
 V(y,x) - \delta(y-x)\int_0^1 V(u,x)\, du\,.
\label{eq:AppB_+}
\end{equation}

The expression for diagram b) requires a more complicated construction
involving $K$:
\begin{equation}
 K(y,x) = C_\text{F}\,\theta(x-y)\frac{1}{x}
         - (x\to\bar{x}, y\to\bar{y})\,.
\end{equation}
The $\log$ terms, containing the ultraviolet scale $\mu_\text{R}^2$
in Table~\ref{tab-5}, are completely cancelled out on account of additional
diagrams with self-energy corrections to the quark legs~\cite{DaCh81},
with contributions of the form
$\ln(\mu_{F}^2/\mu_{R}^2)\cdot C_0(x,\omega)$.
The cancellation of the $\ln(\mu_\text{R}^2)$ terms for the full set of
diagrams is a consequence of the Ward identity in QED.
Finally, collecting the $\log$ terms from all diagrams, one obtains
$\ln(\bar{Q}^2/\mu_{F}^2)\cdot C_0\otimes V_0$ in accordance
with Eq.\ (\ref{eq:factor1}).

To perform the (formal) limit $q^2 \to 0$ in (\ref{eq:factor1}),
one has to take $\omega \to 1$ in the formulas of
Table~\ref{tab-5}, giving rise to the known expression for $T_1$
 \cite{DaCh81,KMR86}, \cite{MNP01a}:
\begin{eqnarray}
 T(Q^2, 0;\mu_\text{F}^2; x)
 &\!\!=\!\!&
 \frac{N_T}{Q^2}
  \left\{\frac1{x}
       + \frac{\alpha_s(\mu_\text{R}^2)}
              {4 \pi}
          \left[\ln\left(\frac{Q^2}{\mu_\text{F}^2}\right)
                 \frac{1}{u}\convo{u}V_0(u,x)
               + \frac{C_\text{F}}{x}
                  \left(\ln^2x
                      - \frac{x\ln x}{1-x}
                      - 9
                  \right)
          \right]
  \right\} \nonumber \\
&&  +\ (x\to \bar{x})
  \ +\ {\cal O}\Big(\frac1{Q^4}\Big)\,;
\label{eq:AppB_NLO}\\
 C_1(Q^2,0;x)
 &=& \frac{C_\text{F}}{x Q^2}
     \left(\ln^2x
         - \frac{x\ln x}{1-x}
         - 9
     \right)
  \ +\ (x\to\bar{x}),
\end{eqnarray}
At the scale $\mu_\text{F}^2=\mu_\text{R}^2=Q^2$, this leads to
\begin{eqnarray}
 \label{eq:AppB_NLO+tw4}
 Q^2 F^{\gamma^*\gamma\pi}(Q^2,0)
  =
  \sqrt{2}f_{\pi}
   \left[1 + \tilde{a}_2 + \tilde{a}_4
       + \frac{\alpha_s(Q^2)C_\text{F}}{4 \pi}
         \left( - 5 + \tilde{a}_2 h_2 + \tilde{a}_4 h_4
         \right)
       - \frac{\delta(Q^2)}{Q^2}\frac{80}{27}
   \right];
\end{eqnarray}
\begin{equation}
 h_n
 \equiv \frac{1}{3} Q^2 C_1(Q^2,0;x)\convo{x}\psi_n(x)\,;
  \quad h_2 = \frac{295}{72}\,;
  \quad h_4 = \frac{10487}{900}\, ,
\label{eq:AppB_NLO+tw4partB}
\end{equation}
where $\tilde{a}_n = a_n^\text{RG}(Q^2)$ and $h_n$ gives the size of the
leading radiative corrections
to the contribution of the $n$th Gegenbauer eigenfunctions
$\psi_n(x)$, entering the expansion of $\varphi_\pi(x)$
(see Eq.\ (\ref{eq:Gegenbauer})).

\subsection{NLO coupling constant}
 \label{App-C2}
\begin{figure}[h]
 \centerline{\includegraphics[width=0.6\textwidth]{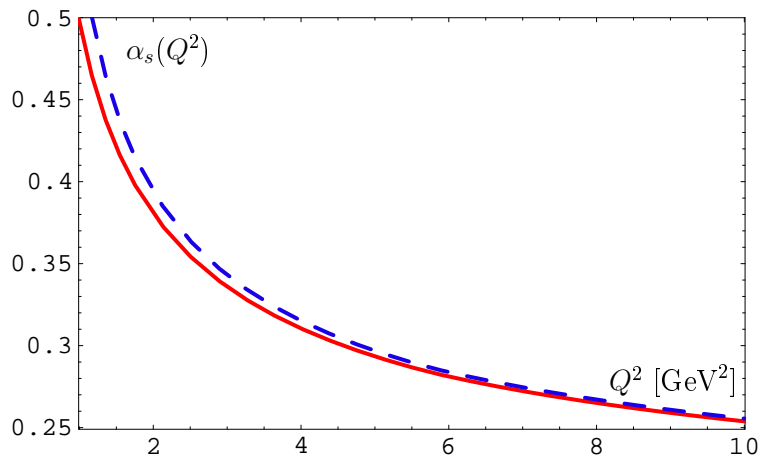}}%
  \caption{\label{fig:als-2E-2B}\footn
    $Q^2$-dependence of two different 2-loop running couplings,
    $\as^\text{2-loop}(Q^2)$ and $\as^\text{PDG}(Q^2)$:
    the solid line corresponds to the exact solution of the RG-equation,
    Eq.\ \protect{(\ref{eq:App-Lambert})};
    the dashed line to the PDG-booklet approximate form,
    Eq.\ \protect{(\ref{eq:PDG-coupling})}.}
\end{figure}

Let us start with the RG equation for the rescaled running coupling
$\Ds a_s\left(Q^2\right)\equiv\as(Q^2)\left(\frac{b_0(N_f)}{4\pi}\right)$:
\begin{equation}
 \label{eq:RG-coupling}
  \frac{d a_s\left(Q^2\right)}{d \ln(Q^2)}
   = \bar{\beta}\left(a_s(Q^2),N_f\right)\, ,
\end{equation}
where $N_f$ is the number of active flavors and the modified
$\bar{\beta}$-function reads
\begin{equation}
 \label{eq:RG-beta}
  \bar{\beta}\left(a_s,N_f\right)
  = -a_s^2
   \left(1
       + c_1 a_s
       + c_2 a_s^2
       + \ldots\
   \right)\, ,
\end{equation}
with $\Ds c_{k}\equiv c_{k}\left(N_f\right)=
     \frac{b_k\left(N_f\right)}{b_0\left(N_f\right)^{k+1}}$,
and the standard $\beta$-function coefficients
are given by
\begin{eqnarray}
 \label{eq:beta-b0}
  b_0\left(N_f\right)
   &=& \frac{11 N_c - 2 N_f}{3}\,;\\
 \label{eq:beta-b1}
  b_1\left(N_f\right)
     &=& \frac{34}{3}N_c^2
       - \left(2C_\text{F} + \frac{10}{3}N_c\right)N_f\,;\\
 \label{eq:beta-b2}
  b_2\left(N_f\right)
     &=& \frac{2857}{2}
       - \frac{5033}{18}N_f
       + \frac{325}{54}N_f^2\, .
\end{eqnarray}
Following here SY, we use the strong coupling constant in
Sec.\ \ref{sec:CLEO} in a ``Particle Data Group'' (PDG) form,
which is the expanded second-order iteration of the 2-loop
equation (\ref{eq:RG-coupling}):
\begin{eqnarray} \label{eq:PDG-coupling}
 \alpha_s^\text{PDG}\left(Q^2, N_f\right) =
   \frac{4\pi}{b_0\left(N_f\right)L(Q^2)}
    \left[ 1 - \frac{L_1(Q^2)}{L(Q^2)}
             +\ \frac{L_1(Q^2)^2
                    - c_1L_1(Q^2)
                    - c_{12}}
                    {L(Q^2)^2}
    \right]\,
\end{eqnarray}
with
\begin{eqnarray}
 L(Q^2)
  \equiv
   \ln\left[\frac{Q^2}{\Lambda\left(N_f\right)^2}
       \right];\,
 L_1(Q^2)
  \equiv
   c_1
    \ln\left[L(Q^2)\right];\,
 c_{12}
  \equiv
   c_{1}^2
   - c_{2}\, ,
\end{eqnarray}
where we fixed $\Lambda_5=\Lambda\left(N_f=5\right)$
with the help of~\cite{PDG2000}
\begin{equation}
 \label{eq:m_Z}
 \alpha_s\left(Q^2=(91.2\ \Gev)^2, N_f=5\right) = 0.118\, .
\end{equation}
Matching this coupling at the $N_f=4$ threshold,
$Q_4 = 10\ \Gev$,
and analogously at the $N_f=3$ threshold,
$Q_3 = 2.4\ \Gev$,
we arrive at
\begin{equation}
  \Lambda_3 = 380\ \Mev\ .
\end{equation}
But one can (we actually did this already in Sec.\ \ref{sec:2loop})
use instead the exact solution of the 2-loop RG equation, rather
than the PDG-booklet expression.
This exact solution can be expressed in terms of the quantity $a_s$
to read
\begin{equation}
 \label{eq:App-asExact}
 \frac{1}{a_s(Q^2)} -
 c_1
     \ln\left[\frac{1}{a_s(Q^2)}+ c_1
     \right] =
 L(Q^2)\, .
\end{equation}
As has been shown in \cite{Mag98} the two-loop running coupling in QCD,
being the solution of this equation, can be written via the Lambert
$W$ function:
\begin{equation}
 \label{eq:App-Lambert}
  \as^\text{2-loop}\left(Q^2,N_f\right)
  = -\frac{4\pi}{b_0\left(N_f\right)c_1}
      \left(1
          + W\left\{-\frac{1}{c_1}
                      \left[\frac{\Lambda^2(N_f)}{e Q^2}
                      \right]^{1/c_1}
             \right\}
      \right)^{-1}\,.
\end{equation}
The difference between the PDG form and the exact solution
varies from 3.5\% at $Q^2=10$~GeV$^2$ to 18\% at $Q^2=1$~GeV$^2$
when one uses the same value of $\Lambda=380$~MeV for both functions.
In a real situation, the values of $\Lambda$
in the two cases are different and are fixed
in accordance with
the standard boundary condition (\ref{eq:m_Z}),
so that $\Lambda_3^\text{PDG}=380$~MeV
and $\Lambda_3^\text{2-loop}=408$~MeV.
The deviation between the two forms becomes less pronounced
at higher $Q^2$
and varies from 0.7\% at $Q^2=10$~GeV$^2$ to 10\% at $Q^2=1$~GeV$^2$
-- see Fig.\ \ref{fig:als-2E-2B}.

It should be realized from this comparison
that the PDG formula (\ref{eq:PDG-coupling})
is afflicted by a large error at $Q^2=1~\gev{2}$
and
for that reason it is preferable in the low $Q^2$-region
($\leq2~\gev{2}$) to use the exact formula (\ref{eq:App-Lambert}).
It is worth to note here 
that in the case of treating the heavy-quark thresholds
in a more accurate way,
as done in~\cite{SM94},
the deviation between the PDG expression 
and the exact formula for the strong coupling 
becomes lower -- at $Q^2=1~\gev{2}$ about 5\%.

\section{The NLO evolution of DA}
 \renewcommand{\theequation}{\thesection.\arabic{equation}}
  \label{App-EvoDA}\setcounter{equation}{0}
The ERBL evolution equation and its kernel have,
respectively, the following structure
(for more details, see, for example,~\cite{Ste99})
\begin{eqnarray}
 \frac{d\,\varphi_{\pi}(x;\mu^2)}
      {d\,\ln\mu^2}
  &=& V(x,u;\alpha_s(\mu^2))\convo{u}\varphi_{\pi}(u;\mu^2)\,;
 \label{eq:App-ERBL}\\
 V(x,u;\alpha_s)
  &=& \left(\frac{\alpha_s}{4\pi}\right)V_0(x,y)
  + \left(\frac{\alpha_s}{4\pi}\right)^2V_1(x,y)
  + \ldots
 \label{eq:App-kernel}
\end{eqnarray}
Its eigenvalues $\gamma_{n}(\alpha_s)$ and
eigenfunctions $\Psi_{n}(x;\alpha_s)$ are defined through
\begin{eqnarray}
 V(x,u;\alpha_s)\convo{u}\Psi_{n}(u;\alpha_s)
  &=& \gamma_n(\alpha_s)\Psi_{n}(x;\alpha_s)\,.
  \label{eq:eigen}
\end{eqnarray}
We use the following notations for its eigenvalues~\cite{Ynd83}
\begin{eqnarray}
 \gamma_n(\alpha_s)
  &=& -\frac{1}{2}
      \left[\left(\frac{\alpha_s}{4\pi}\right)\gamma_0(n)
          + \left(\frac{\alpha_s}{4\pi}\right)^2\gamma_1(n)
          + \ldots
      \right]\,,\label{eq:gamman}
\end{eqnarray}
where the sign ``$-$'' just allows one
to work with positive numbers $\gamma_0(n)$ and $\gamma_1(n)$\footnote{%
The overall factor $1/2$ is due to historical reasons in Eq.(\protect{\ref{eq:gamman}}),
 it is absent in the notations of the original paper~\cite{MR86},
whereas the factor $(-1/2)$ is absent in Muller notations~\cite{Mul94}.}:
\begin{eqnarray}
 \gamma_0(0) &=& 0, \quad \quad \quad
 \gamma_1(0)\ =\ 0\,,
 \\
 \gamma_0(2) &=& \frac{100}{9}, \quad \quad
 \gamma_1(2)\ =\ \frac{34450}{243}-\frac{830}{81}N_f\,,
 \\
 \gamma_0(4) &=& \frac{728}{45}, \quad \quad
 \gamma_1(4)\ =\ \frac{662846}{3375}-\frac{31132}{2025}N_f\, .
\end{eqnarray}
The two-loop eigenfunctions $\Psi_{n}^\text{2-loop}(x;\alpha_s)$
of the ERBL equation (\ref{eq:App-ERBL})
can be expanded
in terms of the one-loop eigenfunctions
$\psi_n(x)$
(see Eq.\ (\ref{eq:Gegenbauer})).
The approximate solution (\ref{eq:EvPDA}) in NLO is then given by
\begin{eqnarray}
 \label{eq:DA_EVO}
  \varphi_\pi^\text{RG}(x,\mu^2)
   &=& \sum\limits_{n} a_{n}(\mu_0^2)E(n,\mu^2)
 \left[\psi_{n}(x) + \frac{\alpha_{s}(\mu)}{4 \pi}
 \sum\limits_{k>n} d_{nk}(\mu^2,\mu_0^2) \psi_{k}(x)
 \right]\,,\\
   E(n,\mu^2)
    &\equiv&
     \exp \left[\int\limits_{a_{s}(\mu_0^2)}
                           ^{a_{s}(\mu^2)}
      \frac{\gamma_n(4\pi a/b_0)\ d a}{\bar{\beta}(a)}
           \right]\,,
\end{eqnarray}
with $a_0=1$, see, e.g., \cite{MR86}.
The ``diagonal'' part (in the $\{\psi_n\}$ basis) of
(\ref{eq:DA_EVO}), expressed by the standard RG exponent, is the
exact part of this solution, while the ``non-diagonal'' part is
taken in the NLO approximation.
The coefficients $d_{nk}(\mu^2,\mu_0^2)$, corresponding to the
non-diagonal part, fix the mixing of the higher harmonics, $k > n$,
due to the fact that the matrix of the anomalous dimensions is
triangular in the $\{\psi_n \}$ basis.
The exponent $E(n,\mu^2)$ in (\ref{eq:DA_EVO}) can be written explicitly
(the scale $\mu_0^2$ can be fixed at some arbitrary value
$\simeq 1\text{ GeV}^2$) to read
\begin{eqnarray}
 \label{eq:Exp_Evo}
   E(n,Q^2)
   &=&
  \left[\frac{a_s(Q^2)}{a_s(\mu^2_0)}
  \right]^{\frac{\gamma_0(n)}{2b_0}}
  \left[\frac{1+c_1a_s(Q^2)}
            {1+c_1a_s(\mu^2_0)}
  \right]^{\omega(n)},
  \\
  \omega(n)
  & \equiv &
      \frac{\gamma_1(n)b_0-\gamma_0(n)b_1}
           {2b_0b_1}\, .
\end{eqnarray}
Evolving according to (\ref{eq:DA_EVO}), the Gegenbauer coefficients
$a_2, a_4$ change from the scale $\mu^2_0=\mu^2_\text{SY}$\footnote{%
Let us remind the reader that we use the values of $a_2$ and $a_4$
fixed at the scale
$\mu^2_\text{SY}=5.76~\text{GeV}^2$ as an input; this is done in order
to facilitate comparison with the SY results.} to the scale $Q^2$
as follows
\begin{eqnarray}
&&a_2 \to  U_2(Q^2, a_2)
   =
  a_2 E(2,Q^2)
  + \frac{\alpha_s(Q^2)}{4\pi}d_{02}(Q^2,\mu^2_0)\,,\\
&&a_4 \to  U_{4}(Q^2, a_2, a_4)
   =
  a_4 E(4,Q^2)
  + \frac{\alpha_s(Q^2)}{4\pi}
  \left[d_{04}(Q^2,\mu^2_0)\right.
\nonumber \\
&&   \phantom{a_4 \to  U_{4}(Q^2, a_2, a_4)=} + \left. a_2 E(2,Q^2)
     d_{24}(Q^2,\mu^2_0)
  \right].
\end{eqnarray}
Here the NLO mixing coefficients are
($k=2, 4\geq n=0, 2$)
\begin{eqnarray}
 d_{nk}(Q^2,\mu^2)
  & = &
  \frac{M_{nk}}{\gamma_0(k)-\gamma_0(n)-2b_0}
  \left\{1
    - \left[\frac{\alpha_s(Q^2)}
                 {\alpha_s(\mu^2)}
      \right]^{[\gamma_0(k)-\gamma_0(n)]/(2b_0)-1}
  \right\}\, ,
\end{eqnarray}
where the values of the first few elements
of the matrix $M_{nk}$ are
\begin{eqnarray}
 \label{eq:App-Mnk}
 M_{0 2} = -11.2 + 1.73  N_f,~~
 M_{0 4} = -1.41 + 0.565 N_f, ~~
 M_{2 4} = -22.0 + 1.65  N_f\,.
\end{eqnarray}
Analytic expressions for $M_{nk}$ have been obtained
in~\cite{Mul94}.
Using them
one can estimate that the accuracy of (\ref{eq:App-Mnk})
is of the order of 1\%.

One appreciates from Eq.\ (\ref{eq:DA_EVO})
that the NLO evolution inevitably generates higher harmonics.
Even in our case,
where we have as a starting point only two harmonics
$a_2(\mu_0^2)\neq0$ and $a_4(\mu_0^2)\neq0$,
the evolution to the scale $Q^2$ produces
$a_k(Q^2)\neq0$ for all $k\geq6$.
As one can see from Eq.\ (\ref{eq:DA_EVO}),
for $k\geq6$ these harmonics are of NLO
($a_k(Q^2)\sim \alpha_s(Q^2)$).
For this reason and owing to
the enormous computional efforts needed for this task,
we take into account only the complete NLO evolution
of the first two nontrivial harmonics.

\section{Expressions for the NLO LCSR transition form factor}
 \renewcommand{\theequation}{\thesection.\arabic{equation}}
  \label{App-OurFF}\setcounter{equation}{0}
We employ a similar formalism as that used in \cite{Kho99,SchmYa99}.
However, to make the present investigation self-contained, we provide
explicit expressions,
given also that the formulas provided in \cite{SchmYa99}
are actually incomplete
and only partially contained in
\cite{Kho99}.
All in all, the NLO LCSR transition form factor is
\begin{eqnarray}
 F_{\pi\gamma^{*}\gamma}(M^2, Q^2, a_2, a_4)
  &=& F_{\pi\gamma^{*}\gamma}^\textsc{LO}(M^2, Q^2, a_2, a_4)
   + \frac{\alpha_s(Q^2)C_\text{F}}{2\pi}\
      F_{\pi\gamma^{*}\gamma}^\textsc{NLO}(M^2, Q^2, a_2, a_4)
      \nonumber\\
  &+& F_{\pi\gamma^{*}\gamma}^\textsc{Tw-4}(M^2, Q^2)
\end{eqnarray}
with
\begin{eqnarray*}
 F_{\pi\gamma^{*}\gamma}^\textsc{(N)LO}(M^2, Q^2, a_2, a_4)
 & = &
   \frac{\sqrt{2}f_{\pi}}{3}
   \left[G_{0}^{\textsc{(N)LO}}(Q^2,M^2)
   +  U_{2}(Q^2, a_2)G_{2}^{\textsc{(N)LO}}(Q^2,M^2)\right.
\nonumber \\
&&  \qquad\quad
   +  \left. U_{4}(Q^2, a_2, a_4)G_{4}^{\textsc{(N)LO}}(Q^2,M^2)
    \right]\, ,
\end{eqnarray*}
\begin{equation}
 F_{\pi\gamma^{*}\gamma}^\textsc{Tw-4}(M^2, Q^2)
  =  \frac{\sqrt{2}f_{\pi}}{3}
    \left[\frac{1}{m_\rho^2}
     \int_{0}^{s_0}
      \rho^\textsc{Tw-4}(Q^2, S)e^{\frac{m_\rho^2-S}{M^2}} dS
  + \frac{1}{Q^2}
     H^\textsc{Tw-4}(Q^2)
   \right]\, ,
\end{equation}
where the evolution functions $U_{2}(Q^2, a_2)$ and
$U_{4}(Q^2, a_2, a_4)$ are described in Appendix \ref{App-EvoDA}.
We also define the following LCSR functions
\begin{eqnarray}
 \label{eq:App-rho_LC1}
 Q^2 G_{k}^{\textsc{order}}(Q^2, M^2)
  & = &
  \frac{Q^2}{m_\rho^2}
   \int_{0}^{s_0}
    \rho_{k}^{\textsc{order}}(Q^2, S)e^{\frac{m_\rho^2-S}{M^2}} dS
  + H_{k}^{\textsc{order}}(Q^2)\,;\\
 \label{eq:App-rho_LC2}
 H_{k}^{\textsc{order}}(Q^2)
  & = &
   \int_{s_0}^\infty
    \rho_{k}^{\textsc{order}}(Q^2, S)
    \frac{Q^2 dS}{S}\,;\\
 \label{eq:App-rho_LC3}
 H^{\textsc{Tw-4}}(Q^2)
  & = &
   \int_{s_0}^\infty
    \rho^{\textsc{Tw-4}}(Q^2, S)
    \frac{Q^2 dS}{S}\, ,
\end{eqnarray}
for $k=0, 2, 4$ and set \textsc{order}$=$ LO or NLO.
The spectral densities in LO are \footnote{%
In the last part of this exposition, we use $u$ instead of $Q^2$
in order to make the formulas more compact.}
\begin{eqnarray}
 \rho_{0}^{\textsc{LO}}(u, s)
  & = & \frac{6 u s}{(u + s)^3}\,;
\\
 \rho_{2}^{\textsc{LO}}(u, s)
  & = & \frac{36 u s(u^2 - 3 u s + s^2)}
             {(u + s)^5}\,;
\\
 \rho_{4}^{\textsc{LO}}(u, s)
  & = & \frac{90 u s(u^4 - 10 u^3 s + 20 u^2 s^2
                         - 10 u s^3 + s^4)}{(u + s)^7}
\end{eqnarray}
and in NLO (see \cite{SchmYa99}) ---
\begin{eqnarray}
 \rho_{0}^{\textsc{NLO}}(u, s)
  & = & \frac{- u s}{(u + s)^3}
  \left[15 - \pi^2 + 3\left(\ln\frac{s}{u}\right)^2
  \right]\,;
\\
 \rho_{2}^{\textsc{NLO}}(u, s)
  & = & \frac{- s}{4(u + s)^5}
          \left[\vphantom{\frac{s}{u}}
                25 s^3 - 8(95+3\pi^2)s^2 u + 36(25+2\pi^2)s u^2
                - 12(5+2\pi^2)u^3
          \right.\nonumber\\
  &   &   \left.~~~~~~~~~~~~~~
                + 12(s^2 - 3s u + u^2) u \ln\frac{s}{u}
                   \left(25 + 6 \ln\frac{s}{u}\right)
           \right]\,;
\\
 \rho_{4}^{\textsc{NLO}}(u, s)
 & = & \frac{- s}{10(u + s)^7}
          \left[\vphantom{\frac{s}{u}}
                91 s^5 - 2(5413+75\pi^2)s^4 u + 125(541+12\pi^2)s^3 u^2
          \right.\nonumber\\
  &   &   \left.~~~~~~~~~~~~~~~
                - 100(901+30\pi^2)s^2 u^3 + 150(193+10\pi^2)s u^4
          \right.\nonumber\\
  &   &   \left.~~~~~~~~~~~~~~~
                - 15(109+10\pi^2)u^5
                + 30(s^4 - 10 s^3 u + 20 s^2 u^2 - 10 s u^3 + u^4) u
          \right.\nonumber\\
  &   &   \left.~~~~~~~~~~~~~~~~~~~~~~~~~~~~~~~~~~~~~~~~~~~~~~
                  \times\ln\frac{s}{u}
                   \left(91 + 15 \ln\frac{s}{u}\right)
           \right]\,.
\end{eqnarray}
Contributions from higher states in LO are given by
\begin{eqnarray}
 H_{0}^{\textsc{LO}}(u)
  & = & \frac{3 u^2}{(u + s_0)^2}\,;
\\
 H_{2}^{\textsc{LO}}(u)
  & = & \frac{3 u^2 (u^2 - 8 u s_0 + 6 s_0^2)}
             {(u + s_0)^4}\,;
\\
 H_{4}^{\textsc{LO}}(u)
  & = & \frac{3 u^2 (u^4 - 24 u^3 s_0 + 90 u^2 s_0^2
                         - 80 u s_0^3 + 15 s_0^4)}{(u + s_0)^6}
\end{eqnarray}
and in NLO:
\begin{eqnarray}
 H_{0}^{\textsc{NLO}}(u)
  & = & \frac{-15 {u^2}}{2(u + s_0)^2}
      - \frac{3 s_0}{(u + s_0)}
         \ln\frac{s_0}{u}
      + \frac{3 (2u + s_0) s_0}{2(u + s_0)^2}
         \left[\left(\ln\frac{s_0}{u}\right)^2
            - \frac{\pi^2}{3}
         \right]\nonumber\\
  & - & 3 \left[\LI{2}\left(-\frac{s_0}{u}\right)
                     - \left(1-\ln\frac{s_0}{u}\right)
                        \ln(1+\frac{s_0}{u})
            \right]\,;\\
 H_{2}^{\textsc{NLO}}(u
 )
  & = & \frac{5 {u}}{48(u + s_0)^4}
         \left(59 {u^3} - 352 {u^2}{s_0}
            + 564 {u}{s_0^2} - 72 {s_0}^3
         \right)\nonumber\\
  & + &  \frac{3 s_0}{2(u + s_0)^4}
          \left(12 u^3 + 4u{s_0}^2 + {s_0}^3\right)
           \left[\left(\ln\frac{s_0}{u}\right)^2
                -\frac{\pi^2}{3}
           \right]\nonumber\\
  & + &  \frac{1}{4(u + s_0)^4}
          \left(5 {u^4} + 4 {s_0}
           \left(47 {u^3} - 3 {s_0}
            \left(18 {u^2} + 3 u {s_0} + {s_0}^2\right)
           \right)
          \right)
          \ln\frac{s_0}{u}\nonumber\\
  & - &  3 \left[\LI{2}\left(-\frac{s_0}{u}\right)
                     - \left(1-\ln\frac{s_0}{u}\right)
                        \ln(1+\frac{s_0}{u})
            \right]
        + \frac{5}{4}\ln(1+\frac{u}{s_0})\,;
\end{eqnarray}
\begin{eqnarray}
 H_{4}^{\textsc{NLO}}(u)
   & = & \frac{u}{600(u + s_0)^6}
          \left(\vphantom{\frac{s_0}{u
          }}
                  10487 {u^5} - 149418 {u^4} {s_0}
                + 678285 {u^3} {s_0}^2 - 888520 {u^2} {s_0}^3
          \right.\nonumber\\
  &   &   \left.\vphantom{\frac{s_0}{u}}~~~~~~~~~~~~~~~~
                + 259260 {u} {s_0}^4 - 6300 {s_0}^5
                         \right)
         \nonumber\\
  & + & \frac{3{s_0}}{2(u + {s_0})^6}
         \left(30 {u^5} - 75 {u^4} {s_0} + 100 {u^3}{s_0}^2
             + 6 u {s_0}^4 + {s_0}^5
         \right)
          \left[\left(\ln\frac{s_0}{u}\right)^2
                       - \frac{\pi^2}{3}
          \right]\nonumber\\
  & + & \frac{u}{10(u + {s_0})^6}
         \left(44 {u^5} + 2124 {u^4} {s_0}
             - 7575 {u^3} {s_0}^2 + 8000 {u^2} {s_0}^3
         \right.\nonumber\\
  &   &  \left.~~~~~~~~~~~~~~~~
             - 1425 {u} {s_0}^4 + 30 {s_0}^5
         \right)
          \ln\frac{s_0}{u}
         -\ 3 \left(\ln\frac{s_0}{u}\right)^2
          \nonumber\\
  & - & 3 \left[\LI{2}\left(-\frac{s_0}{u}\right)
                - \left(1-\ln\frac{s_0}{u}\right)
                   \ln(1+\frac{s_0}{u})
          \right]
        + \frac{7}{5}\ln(1+\frac{u}{s_0})\,.
\end{eqnarray}

The twist-four spectral density is \cite{Kho99}
\begin{eqnarray}
 \rho^{\textsc{Tw-4}}(u, s)
  & = & \frac{80\delta^2(u
  )}{3u}
         \frac{2 u^2 s (s - u)}{(u + s)^5}\, ,
\end{eqnarray}
where $\delta^2(u)$ is taken from Eq.\ (\ref{eq:Ev1L})
with $\mu^2=u$.
This produces
\begin{eqnarray}
 H^{\textsc{Tw-4}}(u)
  & = & \frac{80\delta^2(u)}{3u}
  \frac{u^3(2 s_0 - u)}{3(u + s_0)^4}\, .
\end{eqnarray}

\end{appendix}


\end{document}